\documentclass[aps,prb,twocolumn,reprint,superscriptaddress,showpacs,longbibliography]{revtex4-1}

\usepackage{graphicx}
\usepackage{amsmath}

\DeclareMathOperator{\sgn}{sgn}

\begin{document}

\title{Superfluidity and Density Order in a Bilayer Extended Hubbard Model}

\author{Tuomas I. Vanhala}
\affiliation{COMP Centre of Excellence, Department of Applied Physics, Aalto University, FI-00076 Aalto, Finland}
\author{Jildou E. Baarsma}
\affiliation{COMP Centre of Excellence, Department of Applied Physics, Aalto University, FI-00076 Aalto, Finland}
\author{Miikka O. J. Heikkinen}
\affiliation{COMP Centre of Excellence, Department of Applied Physics, Aalto University, FI-00076 Aalto, Finland}
\author{Matthias Troyer}
\affiliation{Theoretische Physik, ETH Zurich, 8093 Zurich, Switzerland}
\author{Ari Harju}
\affiliation{COMP Centre of Excellence, Department of Applied Physics, Aalto University, FI-00076 Aalto, Finland}
\author{P{\"a}ivi T{\"o}rm{\"a}}
\email[]{paivi.torma@aalto.fi}
\affiliation{COMP Centre of Excellence, Department of Applied Physics, Aalto University, FI-00076 Aalto, Finland}

\date{\today}

\begin{abstract}
We use cellular dynamical mean field theory to study the phase diagram of the square lattice bilayer Hubbard model with an interlayer interaction. The layers are populated by two-component fermions, and the densities in both layers and the strength of the interactions are varied. We find that an attractive interlayer interaction can induce a checkerboard density ordered phase and superfluid phases, with either interlayer or intralayer pairing. Remarkably, the latter phase does not require an intralayer interaction to be present: it can be attributed to an induced attractive interaction caused by density fluctuations in the other layer.
\end{abstract}


\maketitle

\section{Introduction}

Layered lattice systems may exhibit interesting physics since they allow effects typical to two and three dimensions to interplay. For instance, the study of superfluidity in layered systems has been an active area of research since the discovery of high temperature superconductivity in the cuprates \cite{RevModPhys.78.17}. Another intriguing phenomenon in layered systems is the condensation of interlayer electron-hole pairs, or excitons, induced by repulsive interactions between electrons \cite{citeulike:9345659}. Ultracold gases provide one more system where various types of layered systems can be realized. There the possibility to control inter- and intralayer tunnelings (hoppings) has been available for some time \cite{RevModPhys.80.885,PhysRevLett.81.3108} but the interactions have been limited to on-site intralayer type. The interesting new development are dipolar atoms and molecules \cite{citeulike:4444939,CondensedMatte,0953-4075-44-19-193001,PhysRevLett.94.160401,Santos2015,Henn2015} that provide long-range dipole-dipole 
forces and thereby bring in the possibility of interlayer interactions. In this article, we predict within the bilayer extended Hubbard model a competition of density order and two types of superfluids: an interlayer one analogous to exciton condensates, and an {\it intralayer} superfluid that is induced by {\it interlayer} interactions.

Motivated by the variety of experimental systems, several theoretical and computational studies have explored the physics of bilayer or other two-band Hubbard models. Superfluidity and possible modes of pairing were investigated using the dynamical cluster approximation with repulsive intralayer interactions \cite{PhysRevB.84.18} and by Monte Carlo simulations with attractive ones \cite{1367-2630-16-1-013004}. The magnetic properties of the repulsive model were explored in \cite{1367-2630163}. The inclusion of an interlayer or interband interaction opens the possibility of exciton formation and condensation, which has been studied using determinantal Monte Carlo simulations \cite{PhysRevB.88.23}, exact diagonalization \cite{PhysRevB.88.035312}, DMFT \cite{2014arXiv1410.5198K,PhysRevB.89.115134} and various other theoretical approaches \cite{PhysRevB.73.235108,PhysRevB.88.235127}.

Here we consider an extended bilayer Hubbard model, where both layers are populated by two (spin) species of fermionic particles. The interaction between the two layers is attractive, whereas, unlike in most earlier studies, interactions within the layers are vanishing or very weak. Our results are relevant for experiments with dipolar atoms in a bilayer optical lattice, where the dipole moments are aligned by an external field such that the interlayer interaction is attractive \cite{PhysRevLett.108.215301,PhysRevLett.108.210401,PhysRevLett.105.215302,citeulike:12634518,PhysRevLett.101.245301}. Especially ${}^{161}$Dy is a promising candidate: Feshbach resonances are observed and using them the $s$-wave interaction can be tuned to counteract the on-site dipolar interactions \cite{PhysRevA.84.053601,PhysRevA.89.020701,PhysRevB.89.174511}. Moreover, by adjusting the interlayer separation the ratio between intra- and interlayer interactions can be tuned so that nearest-neighbor interactions within the layers are negligible. Our results are equally relevant to the case of a repulsive interlayer interaction, since there is an exact particle-hole transformation between the repulsive and attractive case, and thus a connection to, for instance, exciton condensation can be made \cite{citeulike:9345659}.

The most striking feature we find this model to exhibit is an intralayer superfluid phase, where the Cooper pairs are bound by an interaction {\it induced}, via the interlayer attraction, by {\it density fluctuations in the other layer}. This kind of induced superfluidity has previously been proposed by Kuroki and Aoki \cite{PhysRevB.42.2125,PhysRevLett.69.3820,PhysRevLett.72.2947} as a candidate mechanism for superconductivity in strongly correlated electron systems. Using Monte Carlo methods on the lattice and the bosonization technique in the continuum, they show that, in one dimensional systems, the induced interaction can indeed produce slowly decaying intraband pairing correlations. In this article we consider, in two dimensions, the competition between the intralayer superfluid and other types of long range order. In addition to the induced intralayer superfluid, we find a density ordered phase and an interlayer superfluid. All of these phases can be reached for a fixed interaction strength by tuning only the densities of the layers, making this model a very promising one to realize experimentally.

\section{Model and method}

The hopping Hamiltonian of the two-component bilayer square lattice Hubbard model is,
\begin{eqnarray}
H_h=- t \sum_{\sigma,l, \left \langle i,j \right \rangle} c_{\sigma li}^\dagger c_{\sigma lj} - t_\perp \sum_{\sigma,i} (c_{\sigma 1 i}^\dagger c_{\sigma 2 i} + h.c.) \nonumber \\
- \epsilon \sum_{\sigma,i} (n_{\sigma 1 i}-n_{\sigma 2 i}) - \mu \sum_{\sigma,i} (n_{\sigma 1 i}+n_{\sigma 2 i}),
\label{HoppingHamiltonianFormula}
\end{eqnarray}
where $c_{\sigma li}^\dagger$ creates a particle of spin component $\sigma$ at site $i$ of layer $l$, the corresponding density operators are defined by $n_{\sigma l i}=c_{\sigma li}^\dagger c_{\sigma li}$ and $\left\langle i,j \right\rangle$ means nearest neighbour summation within the layers. The external field $\epsilon$ is related to the interlayer polarization $P_i=\left\langle n_{i1}-n_{i2}\right\rangle$ and the chemical potential $\mu$ can be used to tune the total density $\rho_i=\left\langle n_{i1} +n_{i2} \right\rangle/2$, where $n_{il}=n_{il\uparrow}+n_{il\downarrow}$. Note that $\epsilon$ does not need to be a physical field but can be understood as an energy offset or a difference between chemical potentials of the layers. In this article we assume the interlayer hopping $t_\perp$ to be zero, which can be realized experimentally by using a very deep lattice in the perpendicular direction, and take the intralayer hopping as our unit of energy, $t=1$.

The interaction Hamiltonian is written in the particle-hole symmetric form
\begin{eqnarray}
H_{I}=V \sum_{i,\sigma,\sigma '} \left(n_{\sigma 1 i} - \frac{1}{2} \right)\left (n_{\sigma ' 2 i} - \frac{1}{2}\right) \nonumber\\
+ U \sum_{i,l}\left( n_{\uparrow l i} - \frac{1}{2} \right)\left(n_{\downarrow l i} - \frac{1}{2}\right),
\end{eqnarray}
so that half filling corresponds to $\epsilon=\mu=0$. We focus on the case where the onsite interaction $U$ is tuned to zero (e.g.\ using a Feshbach resonance), and set $V=-3$, corresponding to an attractive interaction between the layers. These parameters can be realized in experiment, for instance, with $^{161}$Dy atoms in a bilayer with intralayer lattice spacing $d=225$ nm and the interlayer spacing $d_\perp=d/3$ and intra- and interlayer lattice heights $V_0=7E_R$ and $V­_{0,\perp}=20V_0$, respectively, with $E_R$ the recoil energy (see appendix \ref{hopping_parameter_appendix}).

To study superfluidity in the model, we use cellular dynamical meand field theory (CDMFT) \cite{RevModPhys.68.,RevModPhys.77.,PhysRevLett.87.186401} with the continuous time auxiliary field (CT-AUX) \cite{ContinuousTime,PhysRevB.83.07,0295-5075-82-5-57003,JPSJ-79-064401} impurity solver that is capable of treating general density-density interactions. This method maps the lattice problem onto a finite size cluster that is self-consistently embedded in the lattice, producing a self energy local to the cluster. Correlation effects within the cluster are treated exactly, and thus the quality of the approximation can be controlled by examining the dependence of the results on cluster size.

The smallest possible cluster that includes both intralayer and interlayer pairing correlations includes one site in both layers. By symmetry arguments (see appendix \ref{order_parameter_appendix}), the relevant superfluid order parameters within this cluster are the intralayer pairing order parameters $\Delta_{l}=\left\langle c_{\uparrow l i} c_{\downarrow l i} \right\rangle$ for both layers $l$ and the interlayer pairing order parameter $\Delta_\perp=\left\langle c_{\uparrow 1 i} c_{\downarrow 2 i} \right\rangle=\left\langle c_{\downarrow 1 i} c_{\uparrow 2 i} \right\rangle$. Here we also consider clusters which consist of $L \times L \times 2$ lattice sites and most of our results have been obtained with $L=2$. We treat the system in the Nambu formalism and include the anomalous Green's functions between all orbitals of opposite spins.

It is interesting to note that doing a particle-hole transformation in layer $2$ leaves the Hamiltonian invariant apart from $\epsilon$ and $\mu$ switching roles and a sign change of $V$. In other words, there exists a transformation between the attractive and the repulsive model. In this transformation, the $\Delta_l$ remain invariant and the interlayer order parameter becomes $\Delta_\perp=\sgn(i)\left\langle c_{\uparrow 1 i} c_{\downarrow 2 i}^\dagger \right\rangle$ measuring pairing correlations of excitonic particle-hole-pairs.

\begin{figure}
\includegraphics[width=\columnwidth]{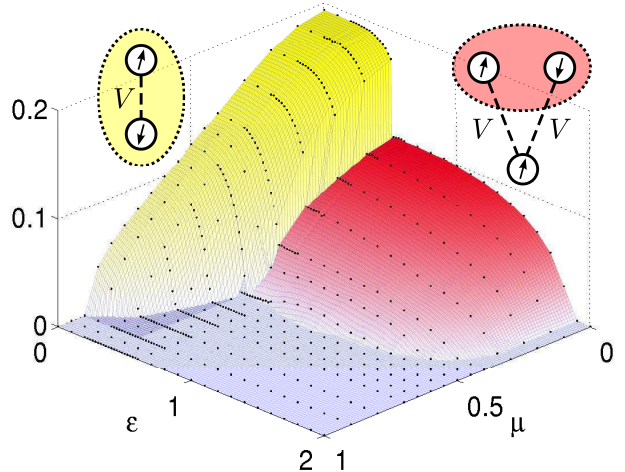}%
\caption{The interlayer order parameter $\Delta_\perp$ (yellow or light grey) and intralayer order parameter $\Delta_l$ for the more dense layer (red or dark grey) from translation invariant two site DMFT, as functions of the interlayer density polarizing field $\epsilon$ and the chemical potential $\mu$. Half filling is given by $\mu=0$. The intralayer order parameter is zero where the interlayer order parameter is nonzero, and vice versa. The inverse temperature is $1/k_BT=15$ and the interaction strength $V=-3$, in units of the hopping $t=1$. The datapoints are marked by black dots, and the surfaces interpolate between them. The Cooper pairs forming the interlayer superfluid are directly bound by the interlayer interaction, while the attraction necessary for the intralayer superfluid is indirectly induced by the other layer. \label{superfluid_figure}}
\end{figure}

\section{Phase diagram}

The intralayer and interlayer superfluid order parameters are shown as a function of the external fields $\mu$ and $\epsilon$ in Fig.~\ref{superfluid_figure}. These results were obtained within two-site DMFT ($L=1$) that excludes translational symmetry breaking. We find that the system prefers interlayer superfluidity for low values of the interlayer density polarizing field $\epsilon$, while higher polarizations drive it into the intralayer superfluid state. The intralayer superfluid region is larger for the more dense layer, i.e. the layer further away from half filling, and therefore the intralayer order parameter is plotted for that layer.

We have thus predicted that, when the symmetry of the layers is distorted by the interlayer density polarizing field, \emph{an intralayer superfluid emerges that does not require intralayer interactions}. An explanation of this intriguing finding could be that the superfluidity is induced by second order effects mediated by the interlayer interaction. To test this hypothesis we calculate the interaction strength in layer $l'$ mediated by density fluctuations in layer $l$ using a simple mean-field theory \cite{Mahan},
\begin{eqnarray}
\nonumber U^{\text{ind}}_{l'}&\propto& 2 V^2\sum_{{\bf k},n} G^0_{{\bf k}nl}G^{0}_{{\bf k}nl}\\
&\propto&\frac{2V^2}{k_BT}\sum_{{\bf k}}n_{\text F}(\varepsilon_{\bf k}-\mu_{l})\left[n_{\text F}(\varepsilon_{\bf k}-\mu_{l})-1\right],
\end{eqnarray}
where in the first line $G^0_{{\bf k}nl}$ is the non-interacting Green's function for the particles in layer $l$ and the summation runs over all energy and momentum states, while the factor 2 results from summing over spin states. In the second line $n_{\text F}$ are the Fermi distribution functions, where $\varepsilon_{\bf k}$ are the particle dipersions and $\mu_l$ determines the density in layer $l$. The strength of $U^{\text{ind}}$ in one layer thus depends on the density in the other layer and turns out to be stronger if mediated by the layer closer to half filling. This explains why the superfluid is more easily formed in the more dense layer, i.e.\ in the layer further away from half filling, while intuition gained from e.g.\ the single band attractive Hubbard model would suggest the opposite.

\begin{figure*}
\includegraphics[width=\textwidth]{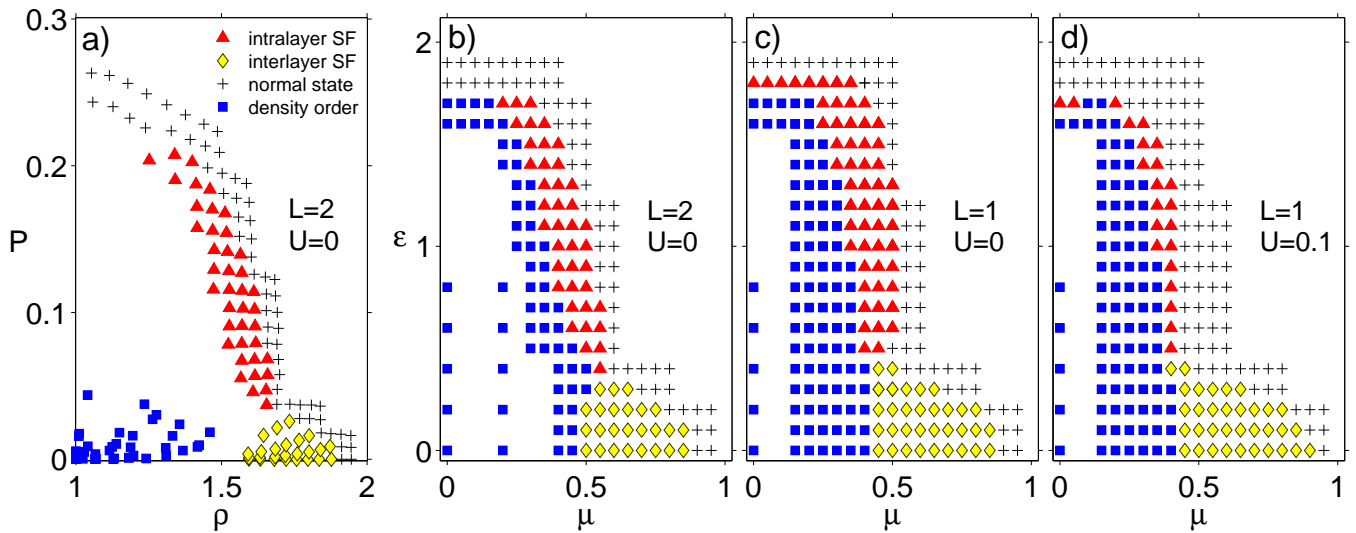}%
\caption{
a) The phase diagram from eigth-site CDMFT at $V=-3$ and $1/k_BT=15$ as a function of total density $\rho$ and polarization $P$. b) and c) Comparison of the $\epsilon$-$\mu$ phase diagrams including density order with cluster sizes $L=1$ and $L=2$. d) The effect of a small intralayer repulsion on the phase diagram.
\label{faasidiagrammit_prb}
}
\end{figure*}

Now, the crucial question is whether the superfluids we predict survive the competition with density order, highly typical for lattice systems. To investigate this, we have obtained the phase diagram of the system also using the $L=2$ cluster, allowing the possibility of a checkerboard density ordered phase, which can be identified by a nonzero value of the order parameter
\begin{equation}
D=\frac{1}{N}\left| \sum_{\sigma l i} \sgn(i) \left\langle n_{\sigma l i}  \right\rangle \right|,
\end{equation}
where $\sgn(i)$ is an alternating sign that has the opposite values for any pair of neighbouring sites inside one layer, and $N$ is the total number of sites on the lattice. The density order is present close to half filling and zero polarization, as can be seen in the phase diagrams in Fig.~\ref{faasidiagrammit_prb}a and \ref{faasidiagrammit_prb}b. Compared to Fig.~\ref{superfluid_figure}, it can be seen that in quite a considerable region where we first found a superfluid phase, actually the density ordered phase turns out to be the ground state of the system. However, we still find a region where the interlayer superfluid phase is the most favourable and, strikingly, the intralayer superfluid phase induced by density fluctuations in the opposite layer remains present in the phase diagram.

Above the critical temperatures of the superfluids, the transition from the density ordered phase to the normal phase is of the second order and gets sharper as the temperature is lowered. At the temperature used in Fig.~\ref{faasidiagrammit_prb} we cannot distinguish this transition from a first order one, which can also be seen in the large gap present in the density - interlayer polarization $\rho-P$ plot. In fact, the competition between the superfluid phases and the density order is expected to cause a true first order transition for almost the whole phase boundary. This is also manifest in the fact that the density ordered region is slightly larger above the superfluid critical temperatures, although it never totally covers the superfluid regions present in Fig.~\ref{faasidiagrammit_prb}. The rapid variation of the density and polarization as a function of $\epsilon$ and $\mu$ suggests phase separation, in particular between the density ordered phase and the intralayer superfluid, for which DMFT solutions can be obtained at the same external polarizing field $\epsilon$ but with large differences in polarization $P$.

\begin{figure}
\includegraphics[width=\columnwidth]{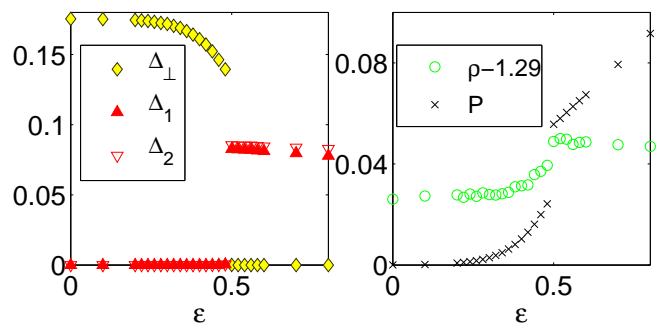}

\caption{The superfluid order parameters (left) and the density $\rho$  and interlayer polarization $P$ (right) from translation invariant two-site DMFT as a function of the polarizing field $\epsilon$ in the intralayer-interlayer transition region. The interaction strength is $V=-3$, the chemical potential $\mu=0.2$ and the inverse temperature $1/k_BT=15$. \label{mu_transition_figure} }

\end{figure}

The translation invariant solution is in itself an interesting model for two band superfluidity where the interband coupling is the dominant one. The order parameters and densities near the intralayer-interlayer transition are shown in Fig.~\ref{mu_transition_figure}. It can be seen that the transition between these two superfluid states is a sharp, first order transition. The interlayer superfluid order parameter $\Delta_\perp$ is zero when the intralayer superfluid order parameter $\Delta_l$ is nonzero and vice versa. The discontinuity in the polarization and the interlayer order parameter gradually decreases with increasing density, and the transition turns into a second order one when the density is high enough so that the intralayer superfluid is not present. Fig.~\ref{mu_transition_figure} can be compared to Fig.~2 of reference \cite{JPSJ-79-064401}, where analogous behaviour is found in a single band attractive Hubbard model as the spin polarization within the single band is increased by a magnetic field. In their case the transition is always a second order one, since the competition between two superfluid phases as in our case is absent.

The intralayer-interlayer superfluid transition is caused by a competition between the different modes of pairing. For instance, if we force the intralayer pairing fields to be zero, the interlayer superfluid region is extended and exhibits a second order transition to the normal state along the whole phase boundary. Similarly, the intralayer superfluid is also present at zero polarization if the interlayer pairing is excluded. Nevertheless, we did not find any hysteretic behaviour in the simulations near the intralayer-interlayer transition, which supports the reliability of the numerics.

\section{Effect of the cluster size and comparison to mean field results}

The critical temperatures of the superfluids and the density ordered phase at selected points in the $(\epsilon,\mu)$ plane are listed in Table \ref{Tc_table}. There is only little variation between different cluster sizes, which is a sign that neglecting nonlocal quantum fluctuations beyond two sites has only a small quantitative effect on the results. This is in contrast to the case of exotic superfluidity in quasi-1D systems \cite{PhysRevLett.113.185301} or d-wave superfluidity in 2D \cite{PhysRevLett.95} where non-local correlations play a crucial role.

It is possible to allow translational symmetry breaking also within the two-site CDMFT by treating the system in a partial real space formalism \cite{PhysRevB.73.205110,1367-2630-10-9-093008} by including two inequivalent impurity problems representing the high-density and low-density sublattices of the checkerboard density order. A comparison of the phase diagrams obtained within this approximation and using the $L=2$ cluster is presented in Fig.~\ref{faasidiagrammit_prb}b and \ref{faasidiagrammit_prb}c. We find that the results are in good agreement, which further supports the conclusion that the effect of the cluster size is not very important.

\begin{figure}
\includegraphics[width=\columnwidth]{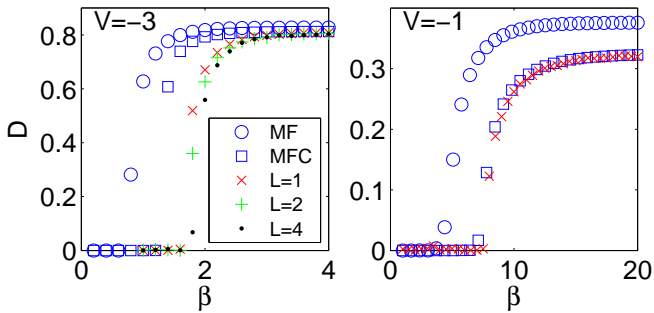}%
\caption{The order parameter $D$ of the density ordered phase as a function of the inverse temperature $\beta=1/k_BT$ from mean field theory (MF), mean field including second order corrections (MFC), and CDMFT with cluster size $L \times L \times 2$. Here, the system is at half filling and zero polarization, $\mu=\epsilon=0$. \label{density_order_transition_figure}}
\end{figure}

The effects of different approximations on the transition from the normal phase to the density order as a function of the inverse temperature are illustrated in Fig.~\ref{density_order_transition_figure}. As expected, the critical temperature is slightly lower for larger clusters, but the variation is small and the effect of the cluster size is insignificant when the system is not very close to the critical point. In contrast, the mean field approximation gives considerably higher critical temperatures, emphasizing the importance of more accurate methods. For a weaker interaction strength $V=-1$ the critical temperature from two-site DMFT ($L=1$) is $T_c=0.13 \pm 0.01$. In this case the system is well in the perturbative region, and while the mean field treatment still gives considerable errors, including second order corrections (see appendix \ref{mean_field_appendix}) already yields a good approximation. For $V=-3$,  however, even with the second order corrections mean field theory is clearly inadequate.

\begin{table}[h]

\begin{center}
\caption{Critical temperatures $T_c$ (in units of $t/k_B$) at selected points in the $(\epsilon,\mu)$ plane for different cluster sizes  $L \times L \times 2$. The interaction strength is $V=-3$.\label{Tc_table} }

    \begin{tabular}{ | l | l | l | l | l |}
    \hline
     Type & Intralayer SF & Interlayer SF & Density Order \\ \hline
    $(\epsilon,\mu)$ & $(1.1,0.45)$ & $(0.0,0.6)$ & $(0.0,0.0)$  \\ \hline
    $T_c$ (L=1) & $0.072 \pm 0.003$ & $0.118 \pm 0.007$ & $0.67 \pm 0.05$ \\ \hline
    $T_c$ (L=2) & $0.072 \pm 0.003$  & $0.112 \pm 0.007$ & $0.63 \pm 0.08$  \\ \hline
    $T_c$ (L=4) &  & & $0.59 \pm 0.04$  \\ \hline
    \end{tabular}

\end{center}

\end{table}

\section{Effect of onsite repulsion and interlayer hopping}

The presence of a small repulsive onsite interaction $U$ does not destroy the intralayer superfluid. For example, the superfluidity persists up to $U \approx 0.1$ at the point $(\epsilon=1.1,\mu=0.4)$, when the other parameters are as in Fig.~\ref{faasidiagrammit_prb}.

The $\mu-\epsilon$ phase diagram for $U=0.1$ is also shown in Fig.~\ref{faasidiagrammit_prb}c. The on-site interaction has only a small effect on the density ordered and interlayer superfluid phases, and the intralayer superfluid also remains present in the phase diagram despite the suppression of intralayer Cooper pairing. On the other hand we find that an attractive onsite interaction (not shown) favours the intralayer superfluid phase as expected.

\begin{figure*}
\includegraphics[width=2\columnwidth]{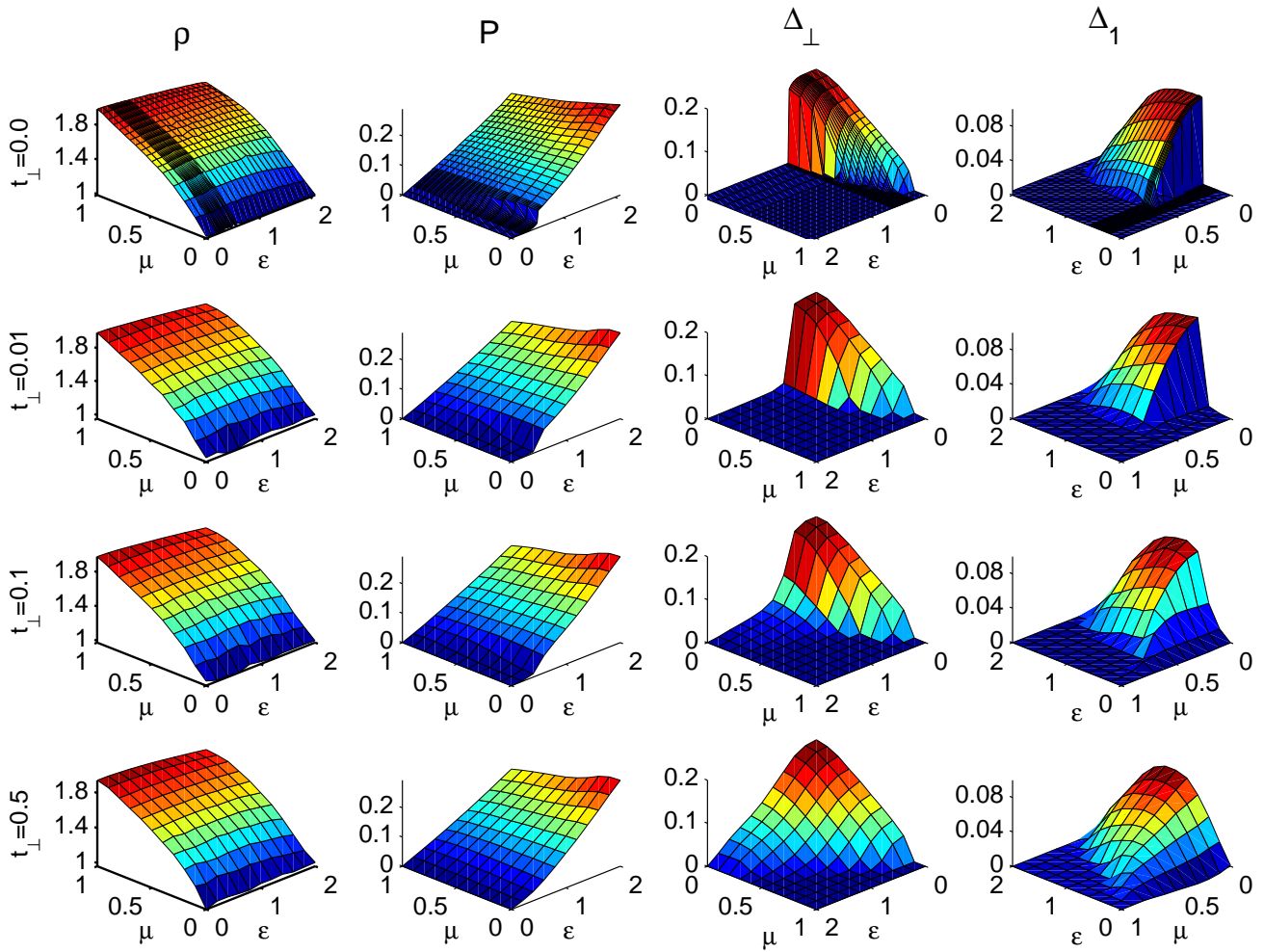}
\caption{A comparison of the superfluid order parameters $\Delta_l$ and $\Delta_\perp$, the density $\rho$ and the polarization $P$ for different values of the interlayer hopping $t_\perp$. These results were obtained within translation invariant two-site DMFT, which excludes the density ordered phase. The parameters were $t=1$, $V=-3$, $U=0$ and the inverse temperature $\beta=15$. Note that in the two rightmost columns the axes $\mu$, $\epsilon$ are rotated to better display the sharpness of the transition. \label{InterlayerHoppingFig}}
\end{figure*}

Finally, we study the effect of a finite interlayer hopping amplitude in the translation invariant case. A comparison of the density, polarization and the superfluid order parameters for different values of $t_\perp$ is shown in Fig.~\ref{InterlayerHoppingFig}. The possibility to tunnel between the layers turns the sharp first order transition between the interlayer and intralayer superfluid states into a smooth crossover. This is expected, since in this case an intralayer Cooper pair can turn into an interlayer one and vice versa thus mixing the two types of superfluids. Note that the smoothening is not very large for $t_\perp=0.1$, and for $t_\perp=0.01$ it is negligible.

\section{Conclusions}

In conclusion, we have studied the bilayer extended Hubbard model for attractive interlayer interactions using a beyond mean field approach. We found density order starting from half filling and, when the density is increased, an interlayer superfluid that appears before reaching the normal state. This superfluid, where the Cooper pairs are formed by particles from different layers, is conceptually related to bilayer exciton condensates where pairing happens between particles in one layer and holes in the other. Notably, the freedom to polarize a density difference between the layers revealed also a novel type of superfluid with intralayer pairing that is induced by density fluctuations in the opposite layer. The predicted phases could be experimentally realized, for instance, with ultracold dipolar atoms or molecules. A prospect for future research is to study if some other type of density order than the checkerboard considered here, possibly incommensurate, would be present near the density order phase boundary. This would be particularly interesting as there is a possibility of simultaneous superfluidity and breaking of translation symmetry, i.e. a supersolid phase.

\begin{acknowledgments}

This work was supported by the Academy of Finland
through its Centers of Excellence Programme (2012-2017)
and under Projects No. 263347, No. 251748, No. 272490,
and by the European Research Council (ERC-2013-AdG-340748-CODE and ERC-2011-AdG-290464-SIMCOFE). T.I.V. is grateful for the support from the Vilho, Yrj\"o and Kalle V\"ais\"al\"a Foundation. Computing resources were
provided by CSC -- the Finnish IT Centre for Science and the
Triton cluster at Aalto University.

\end{acknowledgments}

\appendix

\section{Hopping parameters and energies in the extended Hubbard model}

\label{hopping_parameter_appendix}

In the Hubbard model the hopping parameter $t$ describing hopping from site $j$ to the nearest-neighbor site $j'$ is given by
\begin{align}
t=-\int d^3r\phi_j^*({\bf r})\left[\frac{-\hbar^2}{2m}\nabla^2+V_{latt}({\bf r})\right]\phi_{j'}({\bf r}),
\end{align}
where $V_{latt}({\bf r})$ is the lattice potential. To calculate the interlayer hopping $t_\perp$ in our model, $j$ and $j'$ are taken on different layers, whereas for the intralayer hopping $t$ the sites $j$ and $j'$ are in the same layer. 
The interaction parameters are given by
\begin{align}
\nonumber U&=\frac{4\pi\hbar^2 a_S}{m}\int d^3r|\phi_j({\bf r})|^4\\
&+\int\int d^3rd^3r'|\phi_j({\bf r})|^2|\phi_j({\bf r'})|^2U({\bf r-r'})\\
V_{j,j'}&=\int\int d^3rd^3r'|\phi_j({\bf r})|^2|\phi_{j'}({\bf r'})|^2U({\bf r-r'}),
\end{align}
where $a_S$ is the $s$-wave scattering length and $U({\bf r})$ is the dipole-dipole interaction \cite{Santos2015}, which in the case of all dipoles being aligned reads 
\begin{equation}
U({\bf r})=\frac{C_{dd}}{4\pi}\frac{1-3\cos^2\theta}{r^3},
\end{equation}
where $\theta$ is the angle between ${\bf r}$ and the direction of the polarization.
Here the $C_{dd}$ the strength of the dipolar interaction which in case of dysprosium is $C_{dd}=\mu_0 \mu^2$ with the magnetic dipole moment $\mu=10\mu_B$.
The onsite interaction $U$ results from the contact interaction and the dipole-dipole interaction, whereas the intersite interactions $V_{j,j'}$ result from the dipole-dipole interactions only. 
To make an order of magnitude estimation for the hoppings and interactions we use in this article, we calculated the above coefficients using the tight binding approximation (The onsite wavefunctions $\phi_i({\bf r})$ were taken to be harmonic oscillator ground state wave functions). We found that for $^{161}$Dy in a layer with lattice spacing $d=225$ nm \cite{PhysRevLett.108.215301} and lattice height $V_0=3.5E_R$, with $E_R$ the recoil energy, the onsite interaction can be tuned to zero, $U=0$, since there are several Feshbach resonances available to control $a_S$. An interlayer interaction $V=-1t$ can be reached when the spacing between the layers is $d_\perp=d/3$ and the lattice height in the perpendicular direction is $V­_{0,\perp}=20V_0$. In that case the interlayer hopping $t_\perp$ and intralayer nearest-neighbor interaction $V_\parallel$ are small enough (less than 10\%) in comparison to $t$ and $V$ respectively, to be taken to zero in our model. A stronger interaction, compared to the hopping, can be reached by taking larger lattice heights in all directions. The stronger interaction we consider in this article can be reached when the spacing between the layers is $d_\perp=d/3$ and the lattice heights $V_0=7E_R$ and $V­_{0,\perp}=20V_0$ in the parallel and perpendicular directions, respectively.

\section{Choice of the superfluid order parameters}

\label{order_parameter_appendix}

In this appendix we provide details for the choice of the superfluid order parameters taking into account the symmetries of the system. We first discuss the case $t_\perp=0$ and comment on the general case in the end.

Within the $L=1$ cluster, it is possible to have a six component pairing order parameter which consists of the intralayer order parameter $\Delta_{l}=\left\langle c_{\uparrow l i} c_{\downarrow l i} \right\rangle$ in each layer $l$ and of the four component matrix $M$ defined by
\begin{equation}
\psi_{il}=
	\begin{pmatrix}
		c_{\uparrow l i}   \\
		c_{\downarrow l i}
	\end{pmatrix},
	\quad M=\left\langle \psi_{i1} \psi_{i2}^T \right\rangle.
\end{equation}

As the interlayer hopping vanishes, and the interlayer interaction only couples the total densities of the layers, the Hamiltonian is invariant under rotations $\psi_{il}'=U_l\psi_{il}$ in both layers $l$ where $U_l$ is a unitary matrix. The $SU(2)$-part of this symmetry corresponds to conservation of spin, and the $U(1)$-part corresponds to conservation of particle number, in both layers separately. Letting
\begin{equation}
U_l=
\begin{pmatrix}
	a & b  \\
	c & d
\end{pmatrix},
\end{equation}
the transformation for $\Delta_l$ is
\begin{eqnarray}
\Delta_l' & = \left\langle c_{\uparrow l i}' c_{\downarrow l i}' \right\rangle=\left\langle \left( ac_{\uparrow l i}+bc_{\downarrow l i} \right) \left( cc_{\uparrow l i}+dc_{\downarrow l i} \right) \right\rangle  \nonumber \\
 & = \left\langle (ad-bc) c_{\uparrow l i} c_{\downarrow l i} \right\rangle = \det(U_l)\Delta_l.
\end{eqnarray}
Thus the $\Delta_l$ are invariant in pure $SU(2)$ spin rotations and only gain a phase $\det(U_l)$ in general unitary transformations.

The transformation for the interlayer order parameters is given by
\begin{eqnarray}
M'=\left\langle \psi_{i1}' \left(\psi_{i2}'\right)^T \right\rangle = \left\langle U_1\psi_{i1} \left(U_2\psi_{i2}\right)^T \right\rangle \nonumber \\
= U_1\left\langle \psi_{i1} \left(\psi_{i2}\right)^T \right\rangle U_2^T = U_1 M U_2^T.
\end{eqnarray}
By the singular value decomposition it is always possible to choose $U_1$ and $U_2$ so that $M'$ is diagonal and only has nonnegative real elements. Thus, without losing generality, the order parameter reduces to two nonnegative real valued fields and the two, in general complex valued, $\Delta_{l}$. By doing a further transformation with the matrices
\begin{equation}
U_1=
\begin{pmatrix}
	0 & 1  \\
	1 & 0
\end{pmatrix}, \quad U_2=I,
\end{equation}
$\Delta_1$ gets a minus sign and the interlayer order parameter matrix $M'$ becomes purely off diagonal, the nonzero components being $\left\langle c_{\uparrow 1 i} c_{\downarrow 2 i} \right\rangle$ and $\left\langle c_{\downarrow 1 i} c_{\uparrow 2 i} \right\rangle$.

This is the representation where we perform our numerics. Because of technical limitations we also consider the intralayer order parameters to be real valued. As we have fixed the interlayer order parameters to be nonnegative, this still leaves the signs of the intralayer order parameters as nontrivial factors that cannot be rotated away by symmetry. However, from the simulations we find that the interlayer order parameters (the matrix $M'$) are always zero, when the intralayer order parameters are nonzero, and vice versa. Under this assumption it is always possible to find a transformation that flips the sign of one of the nonzero components, leaving the other one invariant. Thus all sign configurations are equivalent, and we are free to choose the order parameters to be positive, including the intralayer $\Delta_l$. Furthermore, we find that the two nonzero components of $M'$ are always equal, and thus we just call this value the interlayer order parameter $\Delta_\perp \equiv \left\langle c_{\uparrow 1 i} c_{\downarrow 2 i} \right\rangle=\left\langle c_{\downarrow 1 i} c_{\uparrow 2 i} \right\rangle$.

Also within larger clusters, our CDMFT implementation includes all anomalous Green's functions between opposite spin components within the cluster. The $t_\perp=0$ case can be thougth of as a limiting case of a model where the particles have a small probability to tunnel between the layers. This tunneling breaks the conservation of particle number in both layers separately, thus leaving only two exactly conserved spin components. Thus it is expected that the superfluidity can be treated in a picture where the Cooper pairs are only formed between particles of opposite spin, and not between particles of the same spin in different layers. In the case of nonzero $t_\perp$ the intralayer and interlayer order parameters take finite values simultaneously, and the relative signs of the order parameters become important. We find that the preferred sign configuration is such that $\Delta_\perp \equiv \left\langle c_{\uparrow 1 i} c_{\downarrow 2 i} \right\rangle=-\left\langle c_{\downarrow 1 i} c_{\uparrow 2 i} \right\rangle=\left\langle c_{\uparrow 2 i} c_{\downarrow 1 i} \right\rangle$ and $\Delta_2 \Delta_1<0$.

\section{Mean field treatment of the density ordered phase}

\label{mean_field_appendix}

To make comparisons with the CDMFT solution of the model, we have also considered a mean field type approximation for the density ordered phase. The lowest order mean field approximation can be obtained by decomposing the interlayer density-density interaction as
\begin{eqnarray}
& \left(n_A - \frac{1}{2} \right)\left (n_B - \frac{1}{2}\right) \nonumber \\
& = n_An_B-\frac{1}{2} \left(n_A+n_B \right) + \frac{1}{4} \nonumber \\
& = ( \left\langle n_A \right\rangle + \delta_A ) ( \left\langle n_B \right\rangle + \delta_B ) - \frac{1}{2} \left(n_A+n_B \right) + \frac{1}{4} \nonumber \\
& \approx \left\langle n_A \right\rangle \delta_B + \left\langle n_B \right\rangle \delta_A + \left\langle n_A \right\rangle \left\langle n_B \right\rangle - \frac{1}{2} \left(n_A+n_B \right) \nonumber \\
& = \left\langle n_A \right\rangle n_B + \left\langle n_B \right\rangle n_A - \left\langle n_A \right\rangle \left\langle n_B \right\rangle - \frac{1}{2} \left(n_A+n_B \right) \nonumber,
\end{eqnarray}
where $A$ and $B$ stand for spin, site and layer indices, $\delta_A=n_A - \left\langle n_A \right\rangle$ and $\delta_B=n_B - \left\langle n_B \right\rangle$ are fluctuations from the mean value and terms quadratic in the fluctuations have been neglected as well as the constant $1/4$. This leads to the mean field Hamiltonian
\begin{eqnarray}
H=H_h &+ V \sum_{i,\sigma,\sigma ',l} \left( \left\langle n_{\sigma l i} \right\rangle - \frac{1}{2} \right)n_{\sigma ' l' i} \nonumber \\
	  &- V \sum_{i,\sigma,\sigma '} \left\langle n_{\sigma 1 i} \right\rangle \left\langle n_{\sigma ' 2 i} \right\rangle,
\end{eqnarray}
where $l'=1$ when $l=2$ and vice versa. The resulting self energy is diagonal in spin, site and layer indices, and is given in Matsubara frequency space by
\begin{equation}
\Sigma^{(1)}_{\sigma l i}(i\omega_n)=-V \sum_{\sigma'} \left( \left\langle n_{\sigma' l' i} \right\rangle - \frac{1}{2} \right).
\label{MeanFieldSelfEergy}
\end{equation}
Subsequently, the propagator $G_{\sigma l i j}(\tau)=\left\langle c_{\sigma l i}(\tau) c^\dagger_{\sigma l j}(0) \right\rangle$ can be calculated as
\begin{equation}
G_{\sigma l}(\tau)=\frac{1}{\beta}\sum_{\omega_n} \exp\left(-i\omega_n\tau\right)  \left(-i\omega_n + T - \Sigma_{\sigma l}(i\omega_n) \right)^{-1},
\label{PropagatorEquation}
\end{equation}
where $G$, $T$ and $\Sigma$ are matrices in the site indices, $T$ is the hopping matrix of the square lattice including the chemical potential contributions, and $\beta$ is the unitless inverse temperature. In practice the matrix inversion is done in Fourier space with a unit cell that allows the breaking of translation invariance. The self-consistency condition is that the density $\left\langle n_{\sigma l i} \right\rangle = 1-G_{\sigma l i i}(\tau=0+)$ agrees with the density in Eqn. \ref{MeanFieldSelfEergy}.

\begin{figure}
\includegraphics[width=\columnwidth]{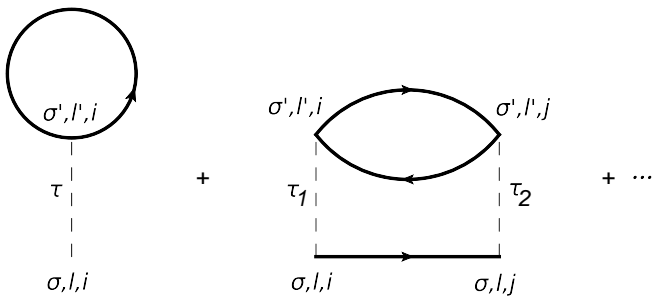}
\caption{The two-particle irreducible diagrams contributing to the self energy up to second order in the interaction strength $V$. The solid lines stand for the full interacting propagator $G$, and the dashed lines represent the interaction vertices. The second order diagram depends on the imaginary time difference $\tau_1-\tau_2$ thus bringing frequency dependence to the self energy. The layer indices must fulfill the condition $l' \neq l$, since there are no interactions within the layers, while the spin index $\sigma'$ can freely be summed over. \label{diagrammit}}
\end{figure}

It is possible to derive the mean field theory from the Baym-Kadanoff (or Luttinger-Ward) functional formalism, which can also be used to systematically include higher order corrections to the approximation \cite{electornicstr,PhysRev.124.287,PhysRev.118.1417}. In this formulation, the self energy is expressed as a diagrammatic series expansion in terms of the interaction vertices and \emph{interacting} propagator lines. As only the two-particle irreducible diagrams enter this series, the contributions up to second order in $V$ consist of only two diagrams, which are depicted in Fig.~\ref{diagrammit}.

Including only the first order diagram gives the mean field self energy \ref{MeanFieldSelfEergy}. The second order diagram includes contributions which are not diagonal in the site indices, see Fig.~\ref{diagrammit}. However, we neglect these contributions and apply a local approximation where both vertices of the diagram are on the same site $i=j$. Evaluating the second order diagram gives the correction term
\begin{eqnarray}
 & \Sigma^{(2)}_{\sigma l i}(\tau_1-\tau_2)= -4V^2 G_{\sigma l i i}(\tau_1-\tau_2)  \cdot \nonumber \\
 & \cdot \sum_{\sigma '} G_{\sigma ' l' i i}(\tau_1-\tau_2) G_{\sigma ' l' i i}(\tau_2-\tau_1),
\end{eqnarray}
where again $l' \neq l$.
We calculate this term directly in an imaginary time grid and then perform a numerical Fourier transformation to Matsubara frequencies. This yields a self-energy 
$\Sigma_{\sigma l i}(i\omega)=\Sigma^{(1)}_{\sigma l i}(i\omega)+\Sigma^{(2)}_{\sigma l i}(i\omega)$ which can be used to calculate the propagator according to Eqn. \ref{PropagatorEquation}. The then obtained propagator is used to calculate the self energy again, and the process is iterated until a converged, self-consistent solution is found.

\bibliography{bilayer_references_clean}

\begin{thebibliography}{45}%
\makeatletter
\providecommand \@ifxundefined [1]{%
 \@ifx{#1\undefined}
}%
\providecommand \@ifnum [1]{%
 \ifnum #1\expandafter \@firstoftwo
 \else \expandafter \@secondoftwo
 \fi
}%
\providecommand \@ifx [1]{%
 \ifx #1\expandafter \@firstoftwo
 \else \expandafter \@secondoftwo
 \fi
}%
\providecommand \natexlab [1]{#1}%
\providecommand \enquote  [1]{``#1''}%
\providecommand \bibnamefont  [1]{#1}%
\providecommand \bibfnamefont [1]{#1}%
\providecommand \citenamefont [1]{#1}%
\providecommand \href@noop [0]{\@secondoftwo}%
\providecommand \href [0]{\begingroup \@sanitize@url \@href}%
\providecommand \@href[1]{\@@startlink{#1}\@@href}%
\providecommand \@@href[1]{\endgroup#1\@@endlink}%
\providecommand \@sanitize@url [0]{\catcode `\\12\catcode `\$12\catcode
  `\&12\catcode `\#12\catcode `\^12\catcode `\_12\catcode `\%12\relax}%
\providecommand \@@startlink[1]{}%
\providecommand \@@endlink[0]{}%
\providecommand \url  [0]{\begingroup\@sanitize@url \@url }%
\providecommand \@url [1]{\endgroup\@href {#1}{\urlprefix }}%
\providecommand \urlprefix  [0]{URL }%
\providecommand \Eprint [0]{\href }%
\providecommand \doibase [0]{http://dx.doi.org/}%
\providecommand \selectlanguage [0]{\@gobble}%
\providecommand \bibinfo  [0]{\@secondoftwo}%
\providecommand \bibfield  [0]{\@secondoftwo}%
\providecommand \translation [1]{[#1]}%
\providecommand \BibitemOpen [0]{}%
\providecommand \bibitemStop [0]{}%
\providecommand \bibitemNoStop [0]{.\EOS\space}%
\providecommand \EOS [0]{\spacefactor3000\relax}%
\providecommand \BibitemShut  [1]{\csname bibitem#1\endcsname}%
\let\auto@bib@innerbib\@empty
\bibitem [{\citenamefont {Lee}\ \emph {et~al.}(2006)\citenamefont {Lee},
  \citenamefont {Nagaosa},\ and\ \citenamefont {Wen}}]{RevModPhys.78.17}%
  \BibitemOpen
  \bibfield  {author} {\bibinfo {author} {\bibfnamefont {Patrick~A.}\
  \bibnamefont {Lee}}, \bibinfo {author} {\bibfnamefont {Naoto}\ \bibnamefont
  {Nagaosa}}, \ and\ \bibinfo {author} {\bibfnamefont {Xiao-Gang}\ \bibnamefont
  {Wen}},\ }\bibfield  {title} {\enquote {\bibinfo {title} {{Doping a Mott
  insulator: Physics of high-temperature superconductivity}},}\ }\href
  {\doibase 10.1103/RevModPhys.78.17} {\bibfield  {journal} {\bibinfo
  {journal} {Rev. Mod. Phys.}\ }\textbf {\bibinfo {volume} {78}},\ \bibinfo
  {pages} {17--85} (\bibinfo {year} {2006})}\BibitemShut {NoStop}%
\bibitem [{\citenamefont {Eisenstein}\ and\ \citenamefont
  {MacDonald}(2004)}]{citeulike:9345659}%
  \BibitemOpen
  \bibfield  {author} {\bibinfo {author} {\bibfnamefont {J.~P.}\ \bibnamefont
  {Eisenstein}}\ and\ \bibinfo {author} {\bibfnamefont {A.~H.}\ \bibnamefont
  {MacDonald}},\ }\bibfield  {title} {\enquote {\bibinfo {title}
  {{Bose{--}Einstein condensation of excitons in bilayer electron systems}},}\
  }\href {\doibase 10.1038/nature03081} {\bibfield  {journal} {\bibinfo
  {journal} {Nature}\ }\textbf {\bibinfo {volume} {432}},\ \bibinfo {pages}
  {691--694} (\bibinfo {year} {2004})}\BibitemShut {NoStop}%
\bibitem [{\citenamefont {Bloch}\ \emph {et~al.}(2008)\citenamefont {Bloch},
  \citenamefont {Dalibard},\ and\ \citenamefont {Zwerger}}]{RevModPhys.80.885}%
  \BibitemOpen
  \bibfield  {author} {\bibinfo {author} {\bibfnamefont {Immanuel}\
  \bibnamefont {Bloch}}, \bibinfo {author} {\bibfnamefont {Jean}\ \bibnamefont
  {Dalibard}}, \ and\ \bibinfo {author} {\bibfnamefont {Wilhelm}\ \bibnamefont
  {Zwerger}},\ }\bibfield  {title} {\enquote {\bibinfo {title} {{Many-body
  physics with ultracold gases}},}\ }\href {\doibase 10.1103/RevModPhys.80.885}
  {\bibfield  {journal} {\bibinfo  {journal} {Rev. Mod. Phys.}\ }\textbf
  {\bibinfo {volume} {80}},\ \bibinfo {pages} {885--964} (\bibinfo {year}
  {2008})}\BibitemShut {NoStop}%
\bibitem [{\citenamefont {Jaksch}\ \emph {et~al.}(1998)\citenamefont {Jaksch},
  \citenamefont {Bruder}, \citenamefont {Cirac}, \citenamefont {Gardiner},\
  and\ \citenamefont {Zoller}}]{PhysRevLett.81.3108}%
  \BibitemOpen
  \bibfield  {author} {\bibinfo {author} {\bibfnamefont {D.}~\bibnamefont
  {Jaksch}}, \bibinfo {author} {\bibfnamefont {C.}~\bibnamefont {Bruder}},
  \bibinfo {author} {\bibfnamefont {J.~I.}\ \bibnamefont {Cirac}}, \bibinfo
  {author} {\bibfnamefont {C.~W.}\ \bibnamefont {Gardiner}}, \ and\ \bibinfo
  {author} {\bibfnamefont {P.}~\bibnamefont {Zoller}},\ }\bibfield  {title}
  {\enquote {\bibinfo {title} {{Cold Bosonic Atoms in Optical Lattices}},}\
  }\href {\doibase 10.1103/PhysRevLett.81.3108} {\bibfield  {journal} {\bibinfo
   {journal} {Phys. Rev. Lett.}\ }\textbf {\bibinfo {volume} {81}},\ \bibinfo
  {pages} {3108--3111} (\bibinfo {year} {1998})}\BibitemShut {NoStop}%
\bibitem [{\citenamefont {Ospelkaus}\ \emph {et~al.}(2008)\citenamefont
  {Ospelkaus}, \citenamefont {Pe{'}er}, \citenamefont {Ni}, \citenamefont
  {Zirbel}, \citenamefont {Neyenhuis}, \citenamefont {Kotochigova},
  \citenamefont {Julienne}, \citenamefont {Ye},\ and\ \citenamefont
  {Jin}}]{citeulike:4444939}%
  \BibitemOpen
  \bibfield  {author} {\bibinfo {author} {\bibfnamefont {S.}~\bibnamefont
  {Ospelkaus}}, \bibinfo {author} {\bibfnamefont {A.}~\bibnamefont {Pe{'}er}},
  \bibinfo {author} {\bibfnamefont {K.~K.}\ \bibnamefont {Ni}}, \bibinfo
  {author} {\bibfnamefont {J.~J.}\ \bibnamefont {Zirbel}}, \bibinfo {author}
  {\bibfnamefont {B.}~\bibnamefont {Neyenhuis}}, \bibinfo {author}
  {\bibfnamefont {S.}~\bibnamefont {Kotochigova}}, \bibinfo {author}
  {\bibfnamefont {P.~S.}\ \bibnamefont {Julienne}}, \bibinfo {author}
  {\bibfnamefont {J.}~\bibnamefont {Ye}}, \ and\ \bibinfo {author}
  {\bibfnamefont {D.~S.}\ \bibnamefont {Jin}},\ }\bibfield  {title} {\enquote
  {\bibinfo {title} {{Efficient state transfer in an ultracold dense gas of
  heteronuclear molecules}},}\ }\href {\doibase 10.1038/nphys997} {\bibfield
  {journal} {\bibinfo  {journal} {Nat Phys}\ }\textbf {\bibinfo {volume} {4}},\
  \bibinfo {pages} {622--626} (\bibinfo {year} {2008})}\BibitemShut {NoStop}%
\bibitem [{\citenamefont {Baranov}\ \emph {et~al.}(2012)\citenamefont
  {Baranov}, \citenamefont {Dalmonte}, \citenamefont {Pupillo},\ and\
  \citenamefont {Zoller}}]{CondensedMatte}%
  \BibitemOpen
  \bibfield  {author} {\bibinfo {author} {\bibfnamefont {M.~A.}\ \bibnamefont
  {Baranov}}, \bibinfo {author} {\bibfnamefont {M.}~\bibnamefont {Dalmonte}},
  \bibinfo {author} {\bibfnamefont {G.}~\bibnamefont {Pupillo}}, \ and\
  \bibinfo {author} {\bibfnamefont {P.}~\bibnamefont {Zoller}},\ }\bibfield
  {title} {\enquote {\bibinfo {title} {{Condensed Matter Theory of Dipolar
  Quantum Gases}},}\ }\href {\doibase 10.1021/cr2003568} {\bibfield  {journal}
  {\bibinfo  {journal} {Chemical Reviews}\ }\textbf {\bibinfo {volume} {112}},\
  \bibinfo {pages} {5012--5061} (\bibinfo {year} {2012})}\BibitemShut {NoStop}%
\bibitem [{\citenamefont {Trefzger}\ \emph {et~al.}(2011)\citenamefont
  {Trefzger}, \citenamefont {Menotti}, \citenamefont {Capogrosso-Sansone},\
  and\ \citenamefont {Lewenstein}}]{0953-4075-44-19-193001}%
  \BibitemOpen
  \bibfield  {author} {\bibinfo {author} {\bibfnamefont {C.}~\bibnamefont
  {Trefzger}}, \bibinfo {author} {\bibfnamefont {C.}~\bibnamefont {Menotti}},
  \bibinfo {author} {\bibfnamefont {B.}~\bibnamefont {Capogrosso-Sansone}}, \
  and\ \bibinfo {author} {\bibfnamefont {M.}~\bibnamefont {Lewenstein}},\
  }\bibfield  {title} {\enquote {\bibinfo {title} {{Ultracold dipolar gases in
  optical lattices}},}\ }\href
  {http://stacks.iop.org/0953-4075/44/i=19/a=193001} {\bibfield  {journal}
  {\bibinfo  {journal} {Journal of Physics B: Atomic, Molecular and Optical
  Physics}\ }\textbf {\bibinfo {volume} {44}},\ \bibinfo {pages} {193001}
  (\bibinfo {year} {2011})}\BibitemShut {NoStop}%
\bibitem [{\citenamefont {Griesmaier}\ \emph {et~al.}(2005)\citenamefont
  {Griesmaier}, \citenamefont {Werner}, \citenamefont {Hensler}, \citenamefont
  {Stuhler},\ and\ \citenamefont {Pfau}}]{PhysRevLett.94.160401}%
  \BibitemOpen
  \bibfield  {author} {\bibinfo {author} {\bibfnamefont {Axel}\ \bibnamefont
  {Griesmaier}}, \bibinfo {author} {\bibfnamefont {J{\"{o}}rg}\ \bibnamefont
  {Werner}}, \bibinfo {author} {\bibfnamefont {Sven}\ \bibnamefont {Hensler}},
  \bibinfo {author} {\bibfnamefont {J{\"{u}}rgen}\ \bibnamefont {Stuhler}}, \
  and\ \bibinfo {author} {\bibfnamefont {Tilman}\ \bibnamefont {Pfau}},\
  }\bibfield  {title} {\enquote {\bibinfo {title} {{Bose-Einstein Condensation
  of Chromium}},}\ }\href {\doibase 10.1103/PhysRevLett.94.160401} {\bibfield
  {journal} {\bibinfo  {journal} {Phys. Rev. Lett.}\ }\textbf {\bibinfo
  {volume} {94}},\ \bibinfo {pages} {160401} (\bibinfo {year}
  {2005})}\BibitemShut {NoStop}%
\bibitem [{\citenamefont {Santos}(2015)}]{Santos2015}%
  \BibitemOpen
  \bibfield  {author} {\bibinfo {author} {\bibfnamefont {L.}~\bibnamefont
  {Santos}},\ }\bibfield  {title} {\enquote {\bibinfo {title} {{Dipolar Gases
  {--} Theory}},}\ }in\ \href@noop {} {\emph {\bibinfo {booktitle} {Quantum Gas
  Experiments {--} Exploring Many-Body States}}},\ \bibinfo {editor} {edited
  by\ \bibinfo {editor} {\bibfnamefont {P.}~\bibnamefont {T{\"{o}}rm{\"{a}}}}\
  and\ \bibinfo {editor} {\bibfnamefont {K.}~\bibnamefont {Sengstock}}}\
  (\bibinfo  {publisher} {Imperial College Press},\ \bibinfo {address}
  {London},\ \bibinfo {year} {2015})\ pp.\ \bibinfo {pages}
  {293--309}\BibitemShut {NoStop}%
\bibitem [{\citenamefont {Henn}\ \emph {et~al.}(2015)\citenamefont {Henn},
  \citenamefont {Billy},\ and\ \citenamefont {Pfau}}]{Henn2015}%
  \BibitemOpen
  \bibfield  {author} {\bibinfo {author} {\bibfnamefont {E.~A.~L.}\
  \bibnamefont {Henn}}, \bibinfo {author} {\bibfnamefont {J.}~\bibnamefont
  {Billy}}, \ and\ \bibinfo {author} {\bibfnamefont {T.}~\bibnamefont {Pfau}},\
  }\bibfield  {title} {\enquote {\bibinfo {title} {{Dipolar Gases {--}
  Experiment}},}\ }in\ \href@noop {} {\emph {\bibinfo {booktitle} {Quantum Gas
  Experiments {--} Exploring Many-Body States}}},\ \bibinfo {editor} {edited
  by\ \bibinfo {editor} {\bibfnamefont {P.}~\bibnamefont {T{\"{o}}rm{\"{a}}}}\
  and\ \bibinfo {editor} {\bibfnamefont {K.}~\bibnamefont {Sengstock}}}\
  (\bibinfo  {publisher} {Imperial College Press},\ \bibinfo {address}
  {London},\ \bibinfo {year} {2015})\ pp.\ \bibinfo {pages}
  {311--325}\BibitemShut {NoStop}%
\bibitem [{\citenamefont {Maier}\ and\ \citenamefont
  {Scalapino}(2011)}]{PhysRevB.84.18}%
  \BibitemOpen
  \bibfield  {author} {\bibinfo {author} {\bibfnamefont {T.~A.}\ \bibnamefont
  {Maier}}\ and\ \bibinfo {author} {\bibfnamefont {D.~J.}\ \bibnamefont
  {Scalapino}},\ }\bibfield  {title} {\enquote {\bibinfo {title} {{Pair
  structure and the pairing interaction in a bilayer Hubbard model for
  unconventional superconductivity}},}\ }\href {\doibase
  10.1103/PhysRevB.84.180513} {\bibfield  {journal} {\bibinfo  {journal} {Phys.
  Rev. B}\ }\textbf {\bibinfo {volume} {84}},\ \bibinfo {pages} {180513}
  (\bibinfo {year} {2011})}\BibitemShut {NoStop}%
\bibitem [{\citenamefont {Zujev}\ \emph {et~al.}(2014)\citenamefont {Zujev},
  \citenamefont {Scalettar}, \citenamefont {Batrouni},\ and\ \citenamefont
  {Sengupta}}]{1367-2630-16-1-013004}%
  \BibitemOpen
  \bibfield  {author} {\bibinfo {author} {\bibfnamefont {Aleksander}\
  \bibnamefont {Zujev}}, \bibinfo {author} {\bibfnamefont {Richard~T.}\
  \bibnamefont {Scalettar}}, \bibinfo {author} {\bibfnamefont {George~G.}\
  \bibnamefont {Batrouni}}, \ and\ \bibinfo {author} {\bibfnamefont {Pinaki}\
  \bibnamefont {Sengupta}},\ }\bibfield  {title} {\enquote {\bibinfo {title}
  {{Pairing correlations in the two-layer attractive Hubbard model}},}\ }\href
  {http://stacks.iop.org/1367-2630/16/i=1/a=013004} {\bibfield  {journal}
  {\bibinfo  {journal} {New Journal of Physics}\ }\textbf {\bibinfo {volume}
  {16}},\ \bibinfo {pages} {013004} (\bibinfo {year} {2014})}\BibitemShut
  {NoStop}%
\bibitem [{\citenamefont {R{\"{u}}ger}\ \emph {et~al.}(2014)\citenamefont
  {R{\"{u}}ger}, \citenamefont {Tocchio}, \citenamefont {Valent{\'{i}}},\ and\
  \citenamefont {Gros}}]{1367-2630163}%
  \BibitemOpen
  \bibfield  {author} {\bibinfo {author} {\bibfnamefont {Robert}\ \bibnamefont
  {R{\"{u}}ger}}, \bibinfo {author} {\bibfnamefont {Luca~F.}\ \bibnamefont
  {Tocchio}}, \bibinfo {author} {\bibfnamefont {Roser}\ \bibnamefont
  {Valent{\'{i}}}}, \ and\ \bibinfo {author} {\bibfnamefont {Claudius}\
  \bibnamefont {Gros}},\ }\bibfield  {title} {\enquote {\bibinfo {title} {{The
  phase diagram of the square lattice bilayer Hubbard model: a variational
  Monte Carlo study}},}\ }\href {\doibase 10.1088/1367-2630/16/3/033010}
  {\bibfield  {journal} {\bibinfo  {journal} {New Journal of Physics}\ }\textbf
  {\bibinfo {volume} {16}},\ \bibinfo {pages} {033010} (\bibinfo {year}
  {2014})}\BibitemShut {NoStop}%
\bibitem [{\citenamefont {Rademaker}\ \emph
  {et~al.}(2013{\natexlab{a}})\citenamefont {Rademaker}, \citenamefont
  {Johnston}, \citenamefont {Zaanen},\ and\ \citenamefont {van~den
  Brink}}]{PhysRevB.88.23}%
  \BibitemOpen
  \bibfield  {author} {\bibinfo {author} {\bibfnamefont {Louk}\ \bibnamefont
  {Rademaker}}, \bibinfo {author} {\bibfnamefont {Steve}\ \bibnamefont
  {Johnston}}, \bibinfo {author} {\bibfnamefont {Jan}\ \bibnamefont {Zaanen}},
  \ and\ \bibinfo {author} {\bibfnamefont {Jeroen}\ \bibnamefont {van~den
  Brink}},\ }\bibfield  {title} {\enquote {\bibinfo {title} {{Determinant
  quantum Monte Carlo study of exciton condensation in the bilayer Hubbard
  model}},}\ }\href {\doibase 10.1103/PhysRevB.88.235115} {\bibfield  {journal}
  {\bibinfo  {journal} {Phys. Rev. B}\ }\textbf {\bibinfo {volume} {88}},\
  \bibinfo {pages} {235115} (\bibinfo {year} {2013}{\natexlab{a}})}\BibitemShut
  {NoStop}%
\bibitem [{\citenamefont {Kaneko}\ \emph {et~al.}(2013)\citenamefont {Kaneko},
  \citenamefont {Ejima}, \citenamefont {Fehske},\ and\ \citenamefont
  {Ohta}}]{PhysRevB.88.035312}%
  \BibitemOpen
  \bibfield  {author} {\bibinfo {author} {\bibfnamefont {T.}~\bibnamefont
  {Kaneko}}, \bibinfo {author} {\bibfnamefont {S.}~\bibnamefont {Ejima}},
  \bibinfo {author} {\bibfnamefont {H.}~\bibnamefont {Fehske}}, \ and\ \bibinfo
  {author} {\bibfnamefont {Y.}~\bibnamefont {Ohta}},\ }\bibfield  {title}
  {\enquote {\bibinfo {title} {{Exact-diagonalization study of exciton
  condensation in electron bilayers}},}\ }\href {\doibase
  10.1103/PhysRevB.88.035312} {\bibfield  {journal} {\bibinfo  {journal} {Phys.
  Rev. B}\ }\textbf {\bibinfo {volume} {88}},\ \bibinfo {pages} {035312}
  (\bibinfo {year} {2013})}\BibitemShut {NoStop}%
\bibitem [{\citenamefont {Kunes}(2014)}]{2014arXiv1410.5198K}%
  \BibitemOpen
  \bibfield  {author} {\bibinfo {author} {\bibfnamefont {J.}~\bibnamefont
  {Kunes}},\ }\bibfield  {title} {\enquote {\bibinfo {title} {{Phase diagram of
  exciton condensate in doped two-band Hubbard model}},}\ }\href@noop {}
  {\bibfield  {journal} {\bibinfo  {journal} {ArXiv e-prints}\ } (\bibinfo
  {year} {2014})},\ \Eprint {http://arxiv.org/abs/1410.5198} {arXiv:1410.5198
  [cond-mat.str-el]} \BibitemShut {NoStop}%
\bibitem [{\citenamefont {Kunes}\ and\ \citenamefont
  {Augustinsk{\'{y}}}(2014)}]{PhysRevB.89.115134}%
  \BibitemOpen
  \bibfield  {author} {\bibinfo {author} {\bibfnamefont {Jan}\ \bibnamefont
  {Kunes}}\ and\ \bibinfo {author} {\bibfnamefont {Pavel}\ \bibnamefont
  {Augustinsk{\'{y}}}},\ }\bibfield  {title} {\enquote {\bibinfo {title}
  {{Excitonic instability at the spin-state transition in the two-band Hubbard
  model}},}\ }\href {\doibase 10.1103/PhysRevB.89.115134} {\bibfield  {journal}
  {\bibinfo  {journal} {Phys. Rev. B}\ }\textbf {\bibinfo {volume} {89}},\
  \bibinfo {pages} {115134} (\bibinfo {year} {2014})}\BibitemShut {NoStop}%
\bibitem [{\citenamefont {Tomio}\ \emph {et~al.}(2006)\citenamefont {Tomio},
  \citenamefont {Honda},\ and\ \citenamefont {Ogawa}}]{PhysRevB.73.235108}%
  \BibitemOpen
  \bibfield  {author} {\bibinfo {author} {\bibfnamefont {Yuh}\ \bibnamefont
  {Tomio}}, \bibinfo {author} {\bibfnamefont {Kotaro}\ \bibnamefont {Honda}}, \
  and\ \bibinfo {author} {\bibfnamefont {Tetsuo}\ \bibnamefont {Ogawa}},\
  }\bibfield  {title} {\enquote {\bibinfo {title} {{Excitonic BCS-BEC crossover
  at finite temperature: Effects of repulsion and electron-hole mass
  difference}},}\ }\href {\doibase 10.1103/PhysRevB.73.235108} {\bibfield
  {journal} {\bibinfo  {journal} {Phys. Rev. B}\ }\textbf {\bibinfo {volume}
  {73}},\ \bibinfo {pages} {235108} (\bibinfo {year} {2006})}\BibitemShut
  {NoStop}%
\bibitem [{\citenamefont {Rademaker}\ \emph
  {et~al.}(2013{\natexlab{b}})\citenamefont {Rademaker}, \citenamefont {van~den
  Brink}, \citenamefont {Zaanen},\ and\ \citenamefont
  {Hilgenkamp}}]{PhysRevB.88.235127}%
  \BibitemOpen
  \bibfield  {author} {\bibinfo {author} {\bibfnamefont {Louk}\ \bibnamefont
  {Rademaker}}, \bibinfo {author} {\bibfnamefont {Jeroen}\ \bibnamefont
  {van~den Brink}}, \bibinfo {author} {\bibfnamefont {Jan}\ \bibnamefont
  {Zaanen}}, \ and\ \bibinfo {author} {\bibfnamefont {Hans}\ \bibnamefont
  {Hilgenkamp}},\ }\bibfield  {title} {\enquote {\bibinfo {title} {{Exciton
  condensation in strongly correlated electron bilayers}},}\ }\href {\doibase
  10.1103/PhysRevB.88.235127} {\bibfield  {journal} {\bibinfo  {journal} {Phys.
  Rev. B}\ }\textbf {\bibinfo {volume} {88}},\ \bibinfo {pages} {235127}
  (\bibinfo {year} {2013}{\natexlab{b}})}\BibitemShut {NoStop}%
\bibitem [{\citenamefont {Lu}\ \emph {et~al.}(2012)\citenamefont {Lu},
  \citenamefont {Burdick},\ and\ \citenamefont {Lev}}]{PhysRevLett.108.215301}%
  \BibitemOpen
  \bibfield  {author} {\bibinfo {author} {\bibfnamefont {Mingwu}\ \bibnamefont
  {Lu}}, \bibinfo {author} {\bibfnamefont {Nathaniel~Q.}\ \bibnamefont
  {Burdick}}, \ and\ \bibinfo {author} {\bibfnamefont {Benjamin~L.}\
  \bibnamefont {Lev}},\ }\bibfield  {title} {\enquote {\bibinfo {title}
  {{Quantum Degenerate Dipolar Fermi Gas}},}\ }\href {\doibase
  10.1103/PhysRevLett.108.215301} {\bibfield  {journal} {\bibinfo  {journal}
  {Phys. Rev. Lett.}\ }\textbf {\bibinfo {volume} {108}},\ \bibinfo {pages}
  {215301} (\bibinfo {year} {2012})}\BibitemShut {NoStop}%
\bibitem [{\citenamefont {Aikawa}\ \emph {et~al.}(2012)\citenamefont {Aikawa},
  \citenamefont {Frisch}, \citenamefont {Mark}, \citenamefont {Baier},
  \citenamefont {Rietzler}, \citenamefont {Grimm},\ and\ \citenamefont
  {Ferlaino}}]{PhysRevLett.108.210401}%
  \BibitemOpen
  \bibfield  {author} {\bibinfo {author} {\bibfnamefont {K.}~\bibnamefont
  {Aikawa}}, \bibinfo {author} {\bibfnamefont {A.}~\bibnamefont {Frisch}},
  \bibinfo {author} {\bibfnamefont {M.}~\bibnamefont {Mark}}, \bibinfo {author}
  {\bibfnamefont {S.}~\bibnamefont {Baier}}, \bibinfo {author} {\bibfnamefont
  {A.}~\bibnamefont {Rietzler}}, \bibinfo {author} {\bibfnamefont
  {R.}~\bibnamefont {Grimm}}, \ and\ \bibinfo {author} {\bibfnamefont
  {F.}~\bibnamefont {Ferlaino}},\ }\bibfield  {title} {\enquote {\bibinfo
  {title} {{Bose-Einstein Condensation of Erbium}},}\ }\href {\doibase
  10.1103/PhysRevLett.108.210401} {\bibfield  {journal} {\bibinfo  {journal}
  {Phys. Rev. Lett.}\ }\textbf {\bibinfo {volume} {108}},\ \bibinfo {pages}
  {210401} (\bibinfo {year} {2012})}\BibitemShut {NoStop}%
\bibitem [{\citenamefont {Pikovski}\ \emph {et~al.}(2010)\citenamefont
  {Pikovski}, \citenamefont {Klawunn}, \citenamefont {Shlyapnikov},\ and\
  \citenamefont {Santos}}]{PhysRevLett.105.215302}%
  \BibitemOpen
  \bibfield  {author} {\bibinfo {author} {\bibfnamefont {A.}~\bibnamefont
  {Pikovski}}, \bibinfo {author} {\bibfnamefont {M.}~\bibnamefont {Klawunn}},
  \bibinfo {author} {\bibfnamefont {G.~V.}\ \bibnamefont {Shlyapnikov}}, \ and\
  \bibinfo {author} {\bibfnamefont {L.}~\bibnamefont {Santos}},\ }\bibfield
  {title} {\enquote {\bibinfo {title} {{Interlayer Superfluidity in Bilayer
  Systems of Fermionic Polar Molecules}},}\ }\href {\doibase
  10.1103/PhysRevLett.105.215302} {\bibfield  {journal} {\bibinfo  {journal}
  {Phys. Rev. Lett.}\ }\textbf {\bibinfo {volume} {105}},\ \bibinfo {pages}
  {215302} (\bibinfo {year} {2010})}\BibitemShut {NoStop}%
\bibitem [{\citenamefont {Yan}\ \emph {et~al.}(2013)\citenamefont {Yan},
  \citenamefont {Moses}, \citenamefont {Gadway}, \citenamefont {Covey},
  \citenamefont {Hazzard}, \citenamefont {Rey}, \citenamefont {Jin},\ and\
  \citenamefont {Ye}}]{citeulike:12634518}%
  \BibitemOpen
  \bibfield  {author} {\bibinfo {author} {\bibfnamefont {Bo}~\bibnamefont
  {Yan}}, \bibinfo {author} {\bibfnamefont {Steven~A.}\ \bibnamefont {Moses}},
  \bibinfo {author} {\bibfnamefont {Bryce}\ \bibnamefont {Gadway}}, \bibinfo
  {author} {\bibfnamefont {Jacob~P.}\ \bibnamefont {Covey}}, \bibinfo {author}
  {\bibfnamefont {Kaden R.~A.}\ \bibnamefont {Hazzard}}, \bibinfo {author}
  {\bibfnamefont {Ana~M.}\ \bibnamefont {Rey}}, \bibinfo {author}
  {\bibfnamefont {Deborah~S.}\ \bibnamefont {Jin}}, \ and\ \bibinfo {author}
  {\bibfnamefont {Jun}\ \bibnamefont {Ye}},\ }\bibfield  {title} {\enquote
  {\bibinfo {title} {{Observation of dipolar spin-exchange interactions with
  lattice-confined polar molecules}},}\ }\href {\doibase 10.1038/nature12483}
  {\bibfield  {journal} {\bibinfo  {journal} {Nature}\ }\textbf {\bibinfo
  {volume} {501}},\ \bibinfo {pages} {521--525} (\bibinfo {year}
  {2013})}\BibitemShut {NoStop}%
\bibitem [{\citenamefont {Bruun}\ and\ \citenamefont
  {Taylor}(2008)}]{PhysRevLett.101.245301}%
  \BibitemOpen
  \bibfield  {author} {\bibinfo {author} {\bibfnamefont {G.~M.}\ \bibnamefont
  {Bruun}}\ and\ \bibinfo {author} {\bibfnamefont {E.}~\bibnamefont {Taylor}},\
  }\bibfield  {title} {\enquote {\bibinfo {title} {{Quantum Phases of a
  Two-Dimensional Dipolar Fermi Gas}},}\ }\href {\doibase
  10.1103/PhysRevLett.101.245301} {\bibfield  {journal} {\bibinfo  {journal}
  {Phys. Rev. Lett.}\ }\textbf {\bibinfo {volume} {101}},\ \bibinfo {pages}
  {245301} (\bibinfo {year} {2008})}\BibitemShut {NoStop}%
\bibitem [{\citenamefont {M{\"{u}}ller}\ \emph {et~al.}(2011)\citenamefont
  {M{\"{u}}ller}, \citenamefont {Billy}, \citenamefont {Henn}, \citenamefont
  {Kadau}, \citenamefont {Griesmaier}, \citenamefont {Jona-Lasinio},
  \citenamefont {Santos},\ and\ \citenamefont {Pfau}}]{PhysRevA.84.053601}%
  \BibitemOpen
  \bibfield  {author} {\bibinfo {author} {\bibfnamefont {S.}~\bibnamefont
  {M{\"{u}}ller}}, \bibinfo {author} {\bibfnamefont {J.}~\bibnamefont {Billy}},
  \bibinfo {author} {\bibfnamefont {E.~A.~L.}\ \bibnamefont {Henn}}, \bibinfo
  {author} {\bibfnamefont {H.}~\bibnamefont {Kadau}}, \bibinfo {author}
  {\bibfnamefont {A.}~\bibnamefont {Griesmaier}}, \bibinfo {author}
  {\bibfnamefont {M.}~\bibnamefont {Jona-Lasinio}}, \bibinfo {author}
  {\bibfnamefont {L.}~\bibnamefont {Santos}}, \ and\ \bibinfo {author}
  {\bibfnamefont {T.}~\bibnamefont {Pfau}},\ }\bibfield  {title} {\enquote
  {\bibinfo {title} {{Stability of a dipolar Bose-Einstein condensate in a
  one-dimensional lattice}},}\ }\href {\doibase 10.1103/PhysRevA.84.053601}
  {\bibfield  {journal} {\bibinfo  {journal} {Phys. Rev. A}\ }\textbf {\bibinfo
  {volume} {84}},\ \bibinfo {pages} {053601} (\bibinfo {year}
  {2011})}\BibitemShut {NoStop}%
\bibitem [{\citenamefont {Baumann}\ \emph {et~al.}(2014)\citenamefont
  {Baumann}, \citenamefont {Burdick}, \citenamefont {Lu},\ and\ \citenamefont
  {Lev}}]{PhysRevA.89.020701}%
  \BibitemOpen
  \bibfield  {author} {\bibinfo {author} {\bibfnamefont {Kristian}\
  \bibnamefont {Baumann}}, \bibinfo {author} {\bibfnamefont {Nathaniel~Q.}\
  \bibnamefont {Burdick}}, \bibinfo {author} {\bibfnamefont {Mingwu}\
  \bibnamefont {Lu}}, \ and\ \bibinfo {author} {\bibfnamefont {Benjamin~L.}\
  \bibnamefont {Lev}},\ }\bibfield  {title} {\enquote {\bibinfo {title}
  {{Observation of low-field Fano-Feshbach resonances in ultracold gases of
  dysprosium}},}\ }\href {\doibase 10.1103/PhysRevA.89.020701} {\bibfield
  {journal} {\bibinfo  {journal} {Phys. Rev. A}\ }\textbf {\bibinfo {volume}
  {89}},\ \bibinfo {pages} {020701} (\bibinfo {year} {2014})}\BibitemShut
  {NoStop}%
\bibitem [{\citenamefont {Zeng}\ and\ \citenamefont
  {Yin}(2014)}]{PhysRevB.89.174511}%
  \BibitemOpen
  \bibfield  {author} {\bibinfo {author} {\bibfnamefont {Tian-Sheng}\
  \bibnamefont {Zeng}}\ and\ \bibinfo {author} {\bibfnamefont {Lan}\
  \bibnamefont {Yin}},\ }\bibfield  {title} {\enquote {\bibinfo {title}
  {{Supersolidity of a dipolar Fermi gas in a cubic optical lattice}},}\ }\href
  {\doibase 10.1103/PhysRevB.89.174511} {\bibfield  {journal} {\bibinfo
  {journal} {Phys. Rev. B}\ }\textbf {\bibinfo {volume} {89}},\ \bibinfo
  {pages} {174511} (\bibinfo {year} {2014})}\BibitemShut {NoStop}%
\bibitem [{\citenamefont {Aoki}\ and\ \citenamefont
  {Kuroki}(1990)}]{PhysRevB.42.2125}%
  \BibitemOpen
  \bibfield  {author} {\bibinfo {author} {\bibfnamefont {Hideo}\ \bibnamefont
  {Aoki}}\ and\ \bibinfo {author} {\bibfnamefont {Kazuhiko}\ \bibnamefont
  {Kuroki}},\ }\bibfield  {title} {\enquote {\bibinfo {title}
  {{Superconductivity in a two-band Hubbard model}},}\ }\href {\doibase
  10.1103/PhysRevB.42.2125} {\bibfield  {journal} {\bibinfo  {journal} {Phys.
  Rev. B}\ }\textbf {\bibinfo {volume} {42}},\ \bibinfo {pages} {2125--2136}
  (\bibinfo {year} {1990})}\BibitemShut {NoStop}%
\bibitem [{\citenamefont {Kuroki}\ and\ \citenamefont
  {Aoki}(1992)}]{PhysRevLett.69.3820}%
  \BibitemOpen
  \bibfield  {author} {\bibinfo {author} {\bibfnamefont {Kazuhiko}\
  \bibnamefont {Kuroki}}\ and\ \bibinfo {author} {\bibfnamefont {Hideo}\
  \bibnamefont {Aoki}},\ }\bibfield  {title} {\enquote {\bibinfo {title}
  {{Realization of negative-U superconductivity in a class of purely repulsive
  systems: Interacting carrier and insulating bands}},}\ }\href {\doibase
  10.1103/PhysRevLett.69.3820} {\bibfield  {journal} {\bibinfo  {journal}
  {Phys. Rev. Lett.}\ }\textbf {\bibinfo {volume} {69}},\ \bibinfo {pages}
  {3820--3823} (\bibinfo {year} {1992})}\BibitemShut {NoStop}%
\bibitem [{\citenamefont {Kuroki}\ and\ \citenamefont
  {Aoki}(1994)}]{PhysRevLett.72.2947}%
  \BibitemOpen
  \bibfield  {author} {\bibinfo {author} {\bibfnamefont {Kazuhiko}\
  \bibnamefont {Kuroki}}\ and\ \bibinfo {author} {\bibfnamefont {Hideo}\
  \bibnamefont {Aoki}},\ }\bibfield  {title} {\enquote {\bibinfo {title}
  {{Superconductivity in a repulsively interacting two-band Fermi gas}},}\
  }\href {\doibase 10.1103/PhysRevLett.72.2947} {\bibfield  {journal} {\bibinfo
   {journal} {Phys. Rev. Lett.}\ }\textbf {\bibinfo {volume} {72}},\ \bibinfo
  {pages} {2947--2950} (\bibinfo {year} {1994})}\BibitemShut {NoStop}%
\bibitem [{\citenamefont {Georges}\ \emph {et~al.}(1996)\citenamefont
  {Georges}, \citenamefont {Kotliar}, \citenamefont {Krauth},\ and\
  \citenamefont {Rozenberg}}]{RevModPhys.68.}%
  \BibitemOpen
  \bibfield  {author} {\bibinfo {author} {\bibfnamefont {Antoine}\ \bibnamefont
  {Georges}}, \bibinfo {author} {\bibfnamefont {Gabriel}\ \bibnamefont
  {Kotliar}}, \bibinfo {author} {\bibfnamefont {Werner}\ \bibnamefont
  {Krauth}}, \ and\ \bibinfo {author} {\bibfnamefont {Marcelo~J.}\ \bibnamefont
  {Rozenberg}},\ }\bibfield  {title} {\enquote {\bibinfo {title} {{Dynamical
  mean-field theory of strongly correlated fermion systems and the limit of
  infinite dimensions}},}\ }\href {\doibase 10.1103/RevModPhys.68.13}
  {\bibfield  {journal} {\bibinfo  {journal} {Rev. Mod. Phys.}\ }\textbf
  {\bibinfo {volume} {68}},\ \bibinfo {pages} {13--125} (\bibinfo {year}
  {1996})}\BibitemShut {NoStop}%
\bibitem [{\citenamefont {Maier}\ \emph
  {et~al.}(2005{\natexlab{a}})\citenamefont {Maier}, \citenamefont {Jarrell},
  \citenamefont {Pruschke},\ and\ \citenamefont {Hettler}}]{RevModPhys.77.}%
  \BibitemOpen
  \bibfield  {author} {\bibinfo {author} {\bibfnamefont {Thomas}\ \bibnamefont
  {Maier}}, \bibinfo {author} {\bibfnamefont {Mark}\ \bibnamefont {Jarrell}},
  \bibinfo {author} {\bibfnamefont {Thomas}\ \bibnamefont {Pruschke}}, \ and\
  \bibinfo {author} {\bibfnamefont {Matthias~H.}\ \bibnamefont {Hettler}},\
  }\bibfield  {title} {\enquote {\bibinfo {title} {{Quantum cluster
  theories}},}\ }\href {\doibase 10.1103/RevModPhys.77.1027} {\bibfield
  {journal} {\bibinfo  {journal} {Rev. Mod. Phys.}\ }\textbf {\bibinfo {volume}
  {77}},\ \bibinfo {pages} {1027--1080} (\bibinfo {year}
  {2005}{\natexlab{a}})}\BibitemShut {NoStop}%
\bibitem [{\citenamefont {Kotliar}\ \emph {et~al.}(2001)\citenamefont
  {Kotliar}, \citenamefont {Savrasov}, \citenamefont {P\'alsson},\ and\
  \citenamefont {Biroli}}]{PhysRevLett.87.186401}%
  \BibitemOpen
  \bibfield  {author} {\bibinfo {author} {\bibfnamefont {Gabriel}\ \bibnamefont
  {Kotliar}}, \bibinfo {author} {\bibfnamefont {Sergej~Y.}\ \bibnamefont
  {Savrasov}}, \bibinfo {author} {\bibfnamefont {Gunnar}\ \bibnamefont
  {P\'alsson}}, \ and\ \bibinfo {author} {\bibfnamefont {Giulio}\ \bibnamefont
  {Biroli}},\ }\bibfield  {title} {\enquote {\bibinfo {title} {{Cellular
  Dynamical Mean Field Approach to Strongly Correlated Systems}},}\ }\href
  {\doibase 10.1103/PhysRevLett.87.186401} {\bibfield  {journal} {\bibinfo
  {journal} {Phys. Rev. Lett.}\ }\textbf {\bibinfo {volume} {87}},\ \bibinfo
  {pages} {186401} (\bibinfo {year} {2001})}\BibitemShut {NoStop}%
\bibitem [{\citenamefont {Gull}\ \emph
  {et~al.}(2011{\natexlab{a}})\citenamefont {Gull}, \citenamefont {Millis},
  \citenamefont {Lichtenstein}, \citenamefont {Rubtsov}, \citenamefont
  {Troyer},\ and\ \citenamefont {Werner}}]{ContinuousTime}%
  \BibitemOpen
  \bibfield  {author} {\bibinfo {author} {\bibfnamefont {E.}~\bibnamefont
  {Gull}}, \bibinfo {author} {\bibfnamefont {A.~J.}\ \bibnamefont {Millis}},
  \bibinfo {author} {\bibfnamefont {A.~I.}\ \bibnamefont {Lichtenstein}},
  \bibinfo {author} {\bibfnamefont {A.~N.}\ \bibnamefont {Rubtsov}}, \bibinfo
  {author} {\bibfnamefont {M.}~\bibnamefont {Troyer}}, \ and\ \bibinfo {author}
  {\bibfnamefont {P.}~\bibnamefont {Werner}},\ }\bibfield  {title} {\enquote
  {\bibinfo {title} {{Continuous-time Monte Carlo methods for quantum impurity
  models}},}\ }\href {\doibase 10.1103/RevModPhys.83.349} {\bibfield  {journal}
  {\bibinfo  {journal} {Reviews of Modern Physics}\ }\textbf {\bibinfo {volume}
  {83}},\ \bibinfo {pages} {349--404} (\bibinfo {year}
  {2011}{\natexlab{a}})}\BibitemShut {NoStop}%
\bibitem [{\citenamefont {Gull}\ \emph
  {et~al.}(2011{\natexlab{b}})\citenamefont {Gull}, \citenamefont {Staar},
  \citenamefont {Fuchs}, \citenamefont {Nukala}, \citenamefont {Summers},
  \citenamefont {Pruschke}, \citenamefont {Schulthess},\ and\ \citenamefont
  {Maier}}]{PhysRevB.83.07}%
  \BibitemOpen
  \bibfield  {author} {\bibinfo {author} {\bibfnamefont {Emanuel}\ \bibnamefont
  {Gull}}, \bibinfo {author} {\bibfnamefont {Peter}\ \bibnamefont {Staar}},
  \bibinfo {author} {\bibfnamefont {Sebastian}\ \bibnamefont {Fuchs}}, \bibinfo
  {author} {\bibfnamefont {Phani}\ \bibnamefont {Nukala}}, \bibinfo {author}
  {\bibfnamefont {Michael~S.}\ \bibnamefont {Summers}}, \bibinfo {author}
  {\bibfnamefont {Thomas}\ \bibnamefont {Pruschke}}, \bibinfo {author}
  {\bibfnamefont {Thomas~C.}\ \bibnamefont {Schulthess}}, \ and\ \bibinfo
  {author} {\bibfnamefont {Thomas}\ \bibnamefont {Maier}},\ }\bibfield  {title}
  {\enquote {\bibinfo {title} {{Submatrix updates for the continuous-time
  auxiliary-field algorithm}},}\ }\href {\doibase 10.1103/PhysRevB.83.075122}
  {\bibfield  {journal} {\bibinfo  {journal} {Phys. Rev. B}\ }\textbf {\bibinfo
  {volume} {83}},\ \bibinfo {pages} {075122} (\bibinfo {year}
  {2011}{\natexlab{b}})}\BibitemShut {NoStop}%
\bibitem [{\citenamefont {Gull}\ \emph {et~al.}(2008)\citenamefont {Gull},
  \citenamefont {Werner}, \citenamefont {Parcollet},\ and\ \citenamefont
  {Troyer}}]{0295-5075-82-5-57003}%
  \BibitemOpen
  \bibfield  {author} {\bibinfo {author} {\bibfnamefont {E.}~\bibnamefont
  {Gull}}, \bibinfo {author} {\bibfnamefont {P.}~\bibnamefont {Werner}},
  \bibinfo {author} {\bibfnamefont {O.}~\bibnamefont {Parcollet}}, \ and\
  \bibinfo {author} {\bibfnamefont {M.}~\bibnamefont {Troyer}},\ }\bibfield
  {title} {\enquote {\bibinfo {title} {{Continuous-time auxiliary-field Monte
  Carlo for quantum impurity models}},}\ }\href
  {http://stacks.iop.org/0295-5075/82/i=5/a=57003} {\bibfield  {journal}
  {\bibinfo  {journal} {EPL (Europhysics Letters)}\ }\textbf {\bibinfo {volume}
  {82}},\ \bibinfo {pages} {57003} (\bibinfo {year} {2008})}\BibitemShut
  {NoStop}%
\bibitem [{\citenamefont {Koga}\ and\ \citenamefont
  {Werner}(2010)}]{JPSJ-79-064401}%
  \BibitemOpen
  \bibfield  {author} {\bibinfo {author} {\bibfnamefont {Akihisa}\ \bibnamefont
  {Koga}}\ and\ \bibinfo {author} {\bibfnamefont {Philipp}\ \bibnamefont
  {Werner}},\ }\bibfield  {title} {\enquote {\bibinfo {title} {{Polarized
  Superfluidity in the Imbalanced Attractive Hubbard Model}},}\ }\href
  {\doibase 10.1143/JPSJ.79.064401} {\bibfield  {journal} {\bibinfo  {journal}
  {Journal of the Physical Society of Japan}\ }\textbf {\bibinfo {volume}
  {79}},\ \bibinfo {pages} {064401} (\bibinfo {year} {2010})}\BibitemShut
  {NoStop}%
\bibitem [{\citenamefont {Mahan}(1993)}]{Mahan}%
  \BibitemOpen
  \bibfield  {author} {\bibinfo {author} {\bibfnamefont {Gerald~D.}\
  \bibnamefont {Mahan}},\ }\href@noop {} {\emph {\bibinfo {title}
  {{Many-Particle Physics}}}},\ \bibinfo {edition} {2nd}\ ed.\ (\bibinfo
  {publisher} {Plenum},\ \bibinfo {address} {New York, N.Y.},\ \bibinfo {year}
  {1993})\BibitemShut {NoStop}%
\bibitem [{\citenamefont {Heikkinen}\ \emph {et~al.}(2014)\citenamefont
  {Heikkinen}, \citenamefont {Kim}, \citenamefont {Troyer},\ and\ \citenamefont
  {T{\"{o}}rm{\"{a}}}}]{PhysRevLett.113.185301}%
  \BibitemOpen
  \bibfield  {author} {\bibinfo {author} {\bibfnamefont {M.~O.~J.}\
  \bibnamefont {Heikkinen}}, \bibinfo {author} {\bibfnamefont {D.~H.}\
  \bibnamefont {Kim}}, \bibinfo {author} {\bibfnamefont {M.}~\bibnamefont
  {Troyer}}, \ and\ \bibinfo {author} {\bibfnamefont {P.}~\bibnamefont
  {T{\"{o}}rm{\"{a}}}},\ }\bibfield  {title} {\enquote {\bibinfo {title}
  {{Nonlocal Quantum Fluctuations and Fermionic Superfluidity in the Imbalanced
  Attractive Hubbard Model}},}\ }\href {\doibase
  10.1103/PhysRevLett.113.185301} {\bibfield  {journal} {\bibinfo  {journal}
  {Phys. Rev. Lett.}\ }\textbf {\bibinfo {volume} {113}},\ \bibinfo {pages}
  {185301} (\bibinfo {year} {2014})}\BibitemShut {NoStop}%
\bibitem [{\citenamefont {Maier}\ \emph
  {et~al.}(2005{\natexlab{b}})\citenamefont {Maier}, \citenamefont {Jarrell},
  \citenamefont {Schulthess}, \citenamefont {Kent},\ and\ \citenamefont
  {White}}]{PhysRevLett.95}%
  \BibitemOpen
  \bibfield  {author} {\bibinfo {author} {\bibfnamefont {T.~A.}\ \bibnamefont
  {Maier}}, \bibinfo {author} {\bibfnamefont {M.}~\bibnamefont {Jarrell}},
  \bibinfo {author} {\bibfnamefont {T.~C.}\ \bibnamefont {Schulthess}},
  \bibinfo {author} {\bibfnamefont {P.~R.~C.}\ \bibnamefont {Kent}}, \ and\
  \bibinfo {author} {\bibfnamefont {J.~B.}\ \bibnamefont {White}},\ }\bibfield
  {title} {\enquote {\bibinfo {title} {{Systematic Study of d-Wave
  Superconductivity in the 2D Repulsive Hubbard Model}},}\ }\href {\doibase
  10.1103/PhysRevLett.95.237001} {\bibfield  {journal} {\bibinfo  {journal}
  {Phys. Rev. Lett.}\ }\textbf {\bibinfo {volume} {95}},\ \bibinfo {pages}
  {237001} (\bibinfo {year} {2005}{\natexlab{b}})}\BibitemShut {NoStop}%
\bibitem [{\citenamefont {Tran}(2006)}]{PhysRevB.73.205110}%
  \BibitemOpen
  \bibfield  {author} {\bibinfo {author} {\bibfnamefont {Minh-Tien}\
  \bibnamefont {Tran}},\ }\bibfield  {title} {\enquote {\bibinfo {title}
  {{Inhomogeneous phases in the Falicov-Kimball model: Dynamical mean-field
  approximation}},}\ }\href {\doibase 10.1103/PhysRevB.73.205110} {\bibfield
  {journal} {\bibinfo  {journal} {Phys. Rev. B}\ }\textbf {\bibinfo {volume}
  {73}},\ \bibinfo {pages} {205110} (\bibinfo {year} {2006})}\BibitemShut
  {NoStop}%
\bibitem [{\citenamefont {Snoek}\ \emph {et~al.}(2008)\citenamefont {Snoek},
  \citenamefont {Titvinidze}, \citenamefont {Tőke}, \citenamefont {Byczuk},\
  and\ \citenamefont {Hofstetter}}]{1367-2630-10-9-093008}%
  \BibitemOpen
  \bibfield  {author} {\bibinfo {author} {\bibfnamefont {M}~\bibnamefont
  {Snoek}}, \bibinfo {author} {\bibfnamefont {I}~\bibnamefont {Titvinidze}},
  \bibinfo {author} {\bibfnamefont {C}~\bibnamefont {Tőke}}, \bibinfo {author}
  {\bibfnamefont {K}~\bibnamefont {Byczuk}}, \ and\ \bibinfo {author}
  {\bibfnamefont {W}~\bibnamefont {Hofstetter}},\ }\bibfield  {title} {\enquote
  {\bibinfo {title} {{Antiferromagnetic order of strongly interacting fermions
  in a trap: real-space dynamical mean-field analysis}},}\ }\href
  {http://stacks.iop.org/1367-2630/10/i=9/a=093008} {\bibfield  {journal}
  {\bibinfo  {journal} {New Journal of Physics}\ }\textbf {\bibinfo {volume}
  {10}},\ \bibinfo {pages} {093008} (\bibinfo {year} {2008})}\BibitemShut
  {NoStop}%
\bibitem [{\citenamefont {Kotliar}\ \emph {et~al.}(2006)\citenamefont
  {Kotliar}, \citenamefont {Savrasov}, \citenamefont {Haule}, \citenamefont
  {Oudovenko}, \citenamefont {Parcollet},\ and\ \citenamefont
  {Marianetti}}]{electornicstr}%
  \BibitemOpen
  \bibfield  {author} {\bibinfo {author} {\bibfnamefont {G.}~\bibnamefont
  {Kotliar}}, \bibinfo {author} {\bibfnamefont {S.~Y.}\ \bibnamefont
  {Savrasov}}, \bibinfo {author} {\bibfnamefont {K.}~\bibnamefont {Haule}},
  \bibinfo {author} {\bibfnamefont {V.~S.}\ \bibnamefont {Oudovenko}}, \bibinfo
  {author} {\bibfnamefont {O.}~\bibnamefont {Parcollet}}, \ and\ \bibinfo
  {author} {\bibfnamefont {C.~A.}\ \bibnamefont {Marianetti}},\ }\bibfield
  {title} {\enquote {\bibinfo {title} {{Electronic structure calculations with
  dynamical mean-field theory}},}\ }\href {\doibase 10.1103/RevModPhys.78.865}
  {\bibfield  {journal} {\bibinfo  {journal} {Reviews of Modern Physics}\
  }\textbf {\bibinfo {volume} {78}},\ \bibinfo {pages} {865--951} (\bibinfo
  {year} {2006})},\ \Eprint {http://arxiv.org/abs/arXiv:cond-mat/0511085}
  {arXiv:cond-mat/0511085} \BibitemShut {NoStop}%
\bibitem [{\citenamefont {Baym}\ and\ \citenamefont
  {Kadanoff}(1961)}]{PhysRev.124.287}%
  \BibitemOpen
  \bibfield  {author} {\bibinfo {author} {\bibfnamefont {Gordon}\ \bibnamefont
  {Baym}}\ and\ \bibinfo {author} {\bibfnamefont {Leo~P.}\ \bibnamefont
  {Kadanoff}},\ }\bibfield  {title} {\enquote {\bibinfo {title} {{Conservation
  Laws and Correlation Functions}},}\ }\href {\doibase 10.1103/PhysRev.124.287}
  {\bibfield  {journal} {\bibinfo  {journal} {Phys. Rev.}\ }\textbf {\bibinfo
  {volume} {124}},\ \bibinfo {pages} {287--299} (\bibinfo {year}
  {1961})}\BibitemShut {NoStop}%
\bibitem [{\citenamefont {Luttinger}\ and\ \citenamefont
  {Ward}(1960)}]{PhysRev.118.1417}%
  \BibitemOpen
  \bibfield  {author} {\bibinfo {author} {\bibfnamefont {J.~M.}\ \bibnamefont
  {Luttinger}}\ and\ \bibinfo {author} {\bibfnamefont {J.~C.}\ \bibnamefont
  {Ward}},\ }\bibfield  {title} {\enquote {\bibinfo {title} {{Ground-State
  Energy of a Many-Fermion System. II}},}\ }\href {\doibase
  10.1103/PhysRev.118.1417} {\bibfield  {journal} {\bibinfo  {journal} {Phys.
  Rev.}\ }\textbf {\bibinfo {volume} {118}},\ \bibinfo {pages} {1417--1427}
  (\bibinfo {year} {1960})}\BibitemShut {NoStop}%
\end{thebibliography}%

\end{document}